\documentclass{pasj01}
\twocolumn

\newcommand\enzo{\texttt{Enzo}}
\newcommand\kpc{\mathrm{\ kpc}}

\newcommand\diff{\mathrm{d}}
\newcommand\HH{\mathrm{H}}

\begin{document}

\title{The Interstellar Medium and Star Formation of Galactic Disks.\\ I. ISM and GMC properties with Diffuse FUV and Cosmic Ray Backgrounds} 

\author{Qi \textsc{Li}\altaffilmark{1}}%
\email{pg3552@ufl.edu}

\author{Jonathan C. \textsc{Tan},\altaffilmark{1,2}}

\author{Duncan \textsc{Christie}\altaffilmark{1}}

\author{Thomas \textsc{G. Bisbas}\altaffilmark{1,4}}

\author{Benjamin \textsc{Wu}\altaffilmark{3}}

\altaffiltext{1}{Department of Astronomy, University of Florida, Gainesville, FL 32611, USA}
\altaffiltext{2}{Department of Physics, University of Florida, Gainesville, FL 32611, USA}
\altaffiltext{3}{National Astronomical Observatory, Mitaka, Tokyo 181-8588, Japan}
\altaffiltext{4}{Max-Planck-Institut f{\"u}r Extraterrestrische Physik, Giessenbachstrasse 1, D-85748 Garching, Germany}


\maketitle
\begin{abstract}
We present a series of adaptive mesh refinement (AMR) hydrodynamic
simulations of flat rotation curve galactic gas disks with a detailed
treatment of the interstellar medium (ISM) physics of the atomic to
molecular phase transition under the influence of diffuse far
ultraviolet (FUV) radiation fields and cosmic ray backgrounds. We
explore the effects of different FUV intensities, including a model
with a radial gradient designed to mimic the Milky Way. The effects of
cosmic rays, including radial gradients in their heating and
ionization rates, are also explored. The final simulations in this
series achieve $4\:$pc resolution across the $\sim20\:$kpc global disk
diameter, with heating and cooling followed down to temperatures of
$\sim10\:$K. The disks are evolved for $300\:$Myr, which is enough
time for the ISM to achieve a quasi-statistical equilibrium. In
particular, the mass fraction of molecular gas stabilizes by $\sim$200
Myr. Additional global ISM properties are analysed. Giant molecular
clouds (GMCs) are also identified and the statistical properties of
their populations examined. GMCs are tracked as the disks evolve. GMC
collisions, which may be a means of triggering star cluster formation,
are counted and the rates compared with analytic models. Relatively
frequent GMC collision rates are seen in these simulations and their
implications for understanding GMC properties, including the driving
of internal turbulence, are discussed.
\end{abstract}

\section{Introduction} 
\label{sec:intro}

Most stars form in galactic disk systems, from ``normal'' disk
galaxies to circumnuclear starburst disks (see, e.g., \cite{KE12};
\cite{Konig16}).  These disks can share a number of common
properties, such as having approximately flat rotation curves (i.e.,
they are shearing disks) and being marginally gravitationally unstable
(i.e., with a \citet{T64} parameter $Q\sim 1$). Observational
evidence has accumulated to connect star formation rates (SFRs) to the global
disk environments (e.g., \cite{K98,Bigiel08,Genzel10,Tan10,Suwannajak14}).
It is thus important to understand the
processes controlling interstellar medium (ISM) dynamics and star
formation activity in such systems.

There have been many previous numerical simulation studies of galactic
disks. The most comparable set of simulations to our current work are
those of Tasker \& Tan (2009, hereafter TT09), \citet{T11} and
\citet{T15}.  TT09 presented a simulation of a galactic disk with total
gas mass surface density ranging from about
$\Sigma_g\simeq60\:M_\odot\:{\rm pc}^{-2}$ at galactocentric radius
$r=2\:$kpc to $\simeq10\:M_\odot\:{\rm pc}^{-2}$ at $10\:$kpc, with
maximum resolution of 8~pc that could undergo cooling to 300~K with a
multi-phase ISM. The cooling floor of 300~K (i.e., a minimum sound
speed of $c_s=1.6\:{\rm km\:s}^{-1}$) was an approximate method of
modeling sub-grid turbulent support. For this reason, the detailed
physics of the atomic to molecular transition was not followed in
these simulations. The TT09 disk fragmented into a population of
self-gravitating giant molecular clouds (GMCs), defined simply as
connected, locally-peaked structures above a threshold density of
$n_{\rm H}=100\:{\rm cm}^{-3}$, which were tracked and had their
collision rates measured. However, this simulation did not include far
ultraviolet (FUV) heating or any other feedback processes. It thus had
a GMC mass fraction that was moderately too high, $\sim 2/3$, compared
to, e.g., the Milky Way disk inside the solar orbit, which has a
molecular mass fraction $f_{\rm mol}\equiv \Sigma_{\rm
  H2}/(\Sigma_{\rm H2}+\Sigma_{\rm HI})\simeq0.75$ at $r=2\:$kpc
falling to $\simeq0.6$ at $5\:$kpc and $\simeq0.2$ by $8\:$kpc
\citep{Koda16}. Also, GMCs in the TT09 simulation had typical mass surface densities,
$\Sigma_{\rm GMC}\sim 300\:M_\odot\:{\rm pc}^{-2}$, i.e., about a factor
of 2 to 3 times higher than observed Milky Way GMCs.

\citet{T11} presented a simulation of the same disk set-up, but now
including a constant diffuse FUV background with intensity of $G_0=4$,
normalized according to \citet{H68}, (i.e., 4 Habings, which is about
2.4 times the local value in the solar vicinity of $G_0=1.7$,
\cite{D78}), which leads to photoelectric heating via dust
grains. A model of star formation, i.e., star particle creation, above
a fixed density threshold of $n_{\rm H,*}=100\:{\rm cm}^{-3}$ at
constant efficiency per local free-fall time \citep{KT07},
but no feedback from young stars. One of the caveats of this
simulation is that identified GMC properties were affected by the
presence of clusters of star particles that could dominate the mass of
the identified gaseous ``GMC'' structure. GMC dynamical properties and
collision rates were thus affected by the presence of these star
clusters. \citet{T15} included localized SNe feedback, in addition to
the diffuse photoelectric heating.  Their result suggests that weak
localized thermal feedback may play a relatively minor role in shaping
the galactic structure compared to gravitational interactions and disk
shear. In both studies, very approximate heating/cooling physics only
down to 300~K was adopted, and the atomic to molecular transition was
not modeled. \citet{Fujimoto14,Fujimoto16}, utilizing the same
simulation code and cooling function as TT09, investigated the GMC
evolution in a M83-type barred spiral galaxy with 1.5 pc
resolution. Frequent cloud-cloud and tidal interactions in the bar
region help to build up massive GMCs and unbound, transient clouds.



With a ``two-fluid'' (isothermal warm and cold gas) model without
thermal processes, \citet{Dobbs08} carried out isolated, magnetized disk
simulations with an SPMHD code and studied GMC formation and evolution
via agglomeration and self gravity in spiral galaxies with different
surface densities. 
\citet{Dobbs11} adopted diffuse FUV heating, radiative and collisional
cooling, and a simple prescription of stellar energy feedback. It was
found that the spiral arms do not significantly trigger star formation
but help gather gas and increase collision rates to produce more
massive and denser GMCs. The GMC mass function was approxiamately
reproduced and similar populations of retrograde and prograde clouds
(relative to the galaxy rotation) found due to enhanced collisions in
the spiral arm regions (although TT09 saw a similar result without
need for large-scale spiral arms). \citet{Dobbs17} extended these
models to study populations of star clusters formed in the disks and
compared them with observed systems. 

Other galactic disk simulations have been conducted to investigate the
physical processes that influence disk evolution. The role of stellar
feedback has been emphasized by, e.g.,
\citet[][]{Agertz13,AK15,Hopkins11,Hopkins12,Hopkins14}.
\citet{Hopkins11, Hopkins12} developed a set of numerical models to
follow stellar feedback on scales from sub-GMC star-forming regions
through entire galaxies, including the energy, momentum, mass, and
metal fluxes from stellar radiation pressure, HII photoionization and
photoelectric heating, Types I and II SNe, and stellar winds (O-star
and AGB).  Based on the models, \citet{Hopkins12,Hopkins13} conducted
pc-resolution SPH simulations of three types of isolated galaxies,
i.e., SMC, Milky Way and Sbc, and produce a quasi-steady ISM in which
GMCs form and disperse rapidly, with phase structure, turbulence, and
disk and GMC properties concluded to be in good agreement with
observations. \citet{Mayer16} studied disk fragmentation and formation
of giant clumps regulated by turbulence via superbubble or blastwave
supernovae feedback, using the GASOLINE2 SPH code and the lagrangian
mesh-less code GIZMO.  The difference in clump properties between the
two sources of turbulence are found as a potential test of feedback
mechanisms.

\citet{GK15,GK16} investigated the driving of turbulence by
gravitational instabilities and star formation feedback, including
stochastic stellar population synthesis, HII region feedback, SNe and
stellar winds via 20-pc-resolution AMR hydrodynamics simulations. They
argue from their results that gravitational instability is likely to
be the dominant source of turbulence and transport in galactic disks,
and that it is responsible for fueling star formation in the inner
parts of galactic disks over cosmological times. The cascade of
turbulence to smaller scales may be one process that regulates the
local star formation rate within GMCs \citep{Padoan02,KM05}, which is
known to occur at a low efficiency per local free-fall time \citep{ZE74,KT07}.


In this work, our overall goal is to understand how the
interstellar medium, including GMCs, and star formation activity in
galactic disk systems is regulated, ultimately modeling all important
physical processes and determining their relative importance,
including in different galactic environments. In this first paper we
introduce our fiducial ``normal'' disk model and our methods of
treating the microphysics of ISM heating and cooling processes. We
start with the simplifying assumption of only considering FUV and CR
heating, which can both be approximated as diffuse components. The
first main goal is to understand ISM structure, including GMC
structural and dynamical properties, in this limiting case, before
complexities of magnetic fields, star formation and localized feedback
are introduced (deferred to future papers in this series). Compared to
the simulations of TT09 the main improvements are: 1) we follow
heating and cooling down to $\sim10\:$K; 2) we use much improved
heating and cooling functions that we develop in this paper based on
photodissociation region (PDR) calculations (for up to four
dimensional grids of density, temperature, FUV intensity and cosmic
ray ionization rate, adopting an empirical extinction versus density
relation to allow local evaluation); 3) a variable mean particle mass
across the atomic to molecular transition is allowed for in the
hydrodynamic equations;
4) we study models, step by step, that investigate the effects of a
variety of different assumptions of the FUV radiation field, including
a model with a radial gradient within the disk; 5) we reach higher
resolutions of 4~pc.


In \S\ref{sec:method} we explain our simulation setup and methods. We
present our results on global ISM properties in \S\ref{sec:ism}. We
examine GMC properties in \S\ref{sec:gmc} and discuss GMC collisions
in \S\ref{S:collisions}. We conclude in \S\ref{S:conclusions}.

\section{Methods} 
\label{sec:method}

\subsection{Numerical Code and Simulation Suite}
\label{sec:nc}

The simulations presented in this paper were run using the numerical
code \enzo, an adaptive mesh refinement (AMR) hydrodynamics code
\citep{B14}. This code solves the hydrodynamics equations using the 3D
\texttt{Zeus} hydrodynamics solver \citep{SN92}, which uses an
artificial viscosity term to handle shocks. The quadratic artificial
viscosity parameter \citep{vN50} was set to 2.0 (the default value)
for all simulations.

For most simulations (Runs I to V), we use a root grid (32.768 kpc on
each side) of $256^3$ and 4 additional AMR levels, giving a minimum
spatial resolution of 8~pc. Cells are refined when the local Jeans
length becomes smaller than 4 cell-widths, which is the condition
typically used to avoid artificial fragmentation \citep{T97}. For
comparison with the results of TT09, in Runs I, II and III, we first
study disks that have a temperature floor of 300~K, which corresponds
to the upper range of temperatures of the atomic cold neutral medium
(CNM) \citep{W03}.  We then adopt a 10 K temperature floor (Run IV, V
and VI) to more accurately follow the atomic to molecular transition
and thus the properties of GMCs. In Run VI we also introduce an extra
level of AMR refinement, i.e., 5 levels in total, achieving 4~pc
resolution to better resolve cloud structures. Note, however, that
with temperatures now followed to 10~K, the Jeans length may not be
well resolved above densities of $\sim$ 400~cm$^{-3}$. Artificial
fragmentation may be occurring above these densities, which should be
kept in mind when interpreting some of the results.

In this paper, we aim to examine the formation and evolution of GMCs
in a flat rotation curve galactic disk without consideration of star
formation, magnetic fields and localized feedback mechanisms. We
mainly study the effects of diffuse FUV feedback and the influence of
diffuse cosmic ray (CR) ionization/heating. To investigate the
influence of diffuse FUV feedback, we present simulation runs with
different static FUV background radiation fields. In Runs I and II we
set up a constant diffuse FUV radiation field with $G_0$ = 1.0, 4.0,
respectively, where $G_0=1$ is the FUV intensity normalized to the
\citet{H68} estimate and a value of $G_0=1.7$ corresponds to the local
FUV radiation field in the Milky Way disk \citep{D78}. In Run III, the
FUV field is set following the profile from \citet{W03}, where we take a local value of $G_0(R_0) = 1.7$ \citep{D78}. 
In Run I to III, the CR
ionization rate is set to a constant value $\zeta = 1\times
10^{-16}\:$s$^{-1}$ (e.g., \cite{Dalgarno06}).  $G_0$ and $\zeta$
are both input parameters of the PDR models.  The configuration of
each run is summarized in Table \ref{tab1}.

\begin{table}[htb!]
\tbl{Configurations of simulations.}{
\begin{tabular}{ccccc}
\hline
Run & Resolution (pc) & $T_\mathrm{floor}$ (K) & $G_0$ & $\zeta$ (${\rm s^{-1}}$)\\
\hline
I & 8 & 300 & 4 & $10^{-16}$\\
II & 8 & 300 & 1 & $10^{-16}$\\
III & 8 & 300 & Wolfire$^a$ & $10^{-16}$\\
IV & 8 & 10 & Wolfire  & $10^{-16}$\\
V & 8 & 10 & Wolfire  & $\zeta(r)$ $^b$\\
VI & 4 & 10 & Wolfire & $10^{-16}$\\
\hline
\end{tabular}}\label{tab1}
\begin{tabnote}
$^a$ FUV intensity radial profile from \citet{W03}.\\
$^b$ Linearly gradient from $\zeta=5\times10^{-16}$ s$^{-1}$ at $r=2\:$kpc to $5\times10^{-17}$~s$^{-1}$ at $r=10\:$kpc.
\end{tabnote}
\end{table}

To better track the atomic to molecular transition, we then carry out
several runs with a 10~K temperature floor. This also permits
investigation of the effects of CR ionization, which mostly influences
the chemistry and heating of cooler, denser (i.e.,
highly-extinguished) molecular gas (i.e., $T<30\:$K, $A_V> 10$~mag,
i.e., $n_{\rm H} > 10^3$ $\mathrm{cm^{-3}}$; see
\cite{Bisbas15}). Note that the temperature floor is not
necessarily reached in the simulations. In Run IV we use a CR ray
ionization rate $\zeta = 1\times 10^{-16}\mathrm{\ s}^{-1}$, while Run
V adopts a simple linear radial profile with the maximum and minimum
equal to $5\times 10^{-16}$ (at $r=2\:$kpc) and $5\times
10^{-17}\:$s$^{-1}$ (at $r=10\:$kpc), respectively. Run VI uses the
same constant value of $\zeta$ as in Run IV, but now introduces an
extra level of AMR to yield 4~pc resolution.

\subsection{Initial Conditions}

Following TT09, we initialize an isolated self-gravitating gas disk,
rotating around the $z$-axis, in a static gravitational background
potential. This potential is described by \citep{BT87}
\begin{equation}
\phi_{\rm bg} = \frac{1}{2} v_{c,0}^2 \ln
    \left[\frac{1}{r_c^2}\left(r_c^2+r^2+\frac{z^2}{q_\phi ^2}\right)\right],
\label{eq:phi}
\end{equation}
i.e., identical to that adopted by TT09, which yields a flat rotation
curve with $v_{c,0}=200\:\mathrm{km\:s^{-1}}$ at galactocentric radii
$r \gg r_c = 0.5\:\mathrm{kpc}$. The axial ratio of the potential
field is $q_\phi=0.7$. The distribution of the background dark matter
and stellar population would in reality evolve in response the
changing distribution of ISM gas mass (e.g., \cite{Read16}),
however, this effect is not included here given our focus on
determining ISM structure that occurs in a given galactic potential.

The initial density profile of the gas disk is set to be
\begin{equation}
\rho(r,z) = \frac{\kappa c_s}{2 \pi G Q z_h} \mathrm{sech^2}\left(\frac{z}{z_h}\right)\:\: (r<12\ \mathrm{kpc}), 
\end{equation}
where the scale height $z_h$ is set to 290~pc, and for purposes of
normalization of the initial condition, the sound speed $c_s$ is set
equal to $7.7\:\rm km\:s^{-1}$. These choices are made so that the
initial \citet{T64} stability parameter, i.e., with input quantities
averaged in annuli in the disk, is $Q=1.5$ for
$2\mathrm{\ kpc}<r<10\mathrm{\ kpc}$, which is our only region of
interest, and $Q=20$ at the same sound speed for $r<2\kpc$ or
$10\kpc<r<12\kpc$, which are boundary regions.  The epicyclic
frequency $\kappa$ is expressed by
\begin{equation}
\kappa = \sqrt{2}\frac{v_{\rm circ}}{r}(1+\beta)^{\frac{1}{2}},
\end{equation}
where $v_{\rm circ}$ is the circular velocity, which is equal to
$v_{c,0}r(r_c^2+r^2)^{-1/2}$ set by the gravitational potential of equation
(\ref{eq:phi}), and $\beta=\diff\ln v_{\rm circ}/\diff\ln r$. In total, the
disk gas mass is $\sim 6 \times 10^9\:M_\odot$.

The disk is surrounded by a quasi-static, dynamically negligible
ambient medium with initial gas density
\begin{equation}
\rho_{\rm ambient} = \rho_0 {\rm exp}{\left(-\frac{\mu m_\mathrm{H}}{kT}\phi\right)}, 
\end{equation}
where the gravitational potential $\phi$ includes the contribution of
background potential and the galactic gas disk. In our simulation, the
maximum value of the initial ambient density is $n_\HH = 10^{-4}\:
\mathrm{cm^{-3}}$ (approximately $10^{-4}$ of that in the main
disk). We set its initial temperature to be $10^6$~K. The ambient
density starts to dominate over that of the main disk component at
several disk scale heights, e.g., at a height of $\sim$1.4 kpc at
$r\sim4$ kpc.

As discussed below, the gas undergoes heating and cooling
processes. The net effect of the evolution from the initial condition
is cooling, which leads to a reduction of $Q$ and thus fragmentation
in the main disk region. Note, we adopt $c_s= \sqrt{\gamma k T / \mu}$
with a fixed adiabatic index $\gamma = 5/3$. In the limit of fully
molecular $\rm H_2$, the mean molecular weight $\mu = 2.33
m_\mathrm{H}$ (i.e., we assume 0.1 He per H and ignore contributions
from other species, including $1\times 10^{-4}$ C and $3\times
10^{-4}$ O per H). The true mean molecular weight $\mu$ varies based
on thermal processes (determined in an iterative way, see
\S\ref{sec:tp}). We adopt the value of $\gamma = 5/3$ for the entire
simulation domain. While this does not account for the excitation of
rotational and vibrational modes of $\rm H_2$ that would occur in
shocks, we consider that this is the most appropriate single-valued
choice of $\gamma$ for our simulation setup, given our focus on the
dynamics of the dense molecular gas and given the resolutions that are
achieved in the models, including allowance for the fact that the
\texttt{Zeus} solver itself does not allow for very accurate shock
capturing.

\subsection{Heating and Cooling Functions}
\label{sec:tp}

To investigate the effects of diffuse FUV heating and CR ionization,
we implement PDR-based heating and cooling functions. The method is
based on that developed by \citet{W15} in the context of simulations of
GMC collisions and for the simple case of a fixed background FUV
intensity ($G_0=4$) and CR ionization rate ($\zeta=10^{-16}\:{\rm
  s}^{-1}$). For $T<10^4\:$K, we use the P{\scriptsize Y}PDR code,
which includes $\sim 30$ species and performs well in producing
results similar to larger PDR codes up to $\sim10^4\:$K
\citep{R07}. P{\scriptsize Y}PDR is also more flexible in being able
to be adapted to calculate self-consistent grids of models at
non-equilibrium conditions, including molecular self-shielding effects
(see \cite{W15}). For $T>10^4\:$K, we utilize the C{\scriptsize
  LOUDY} code \citep{F13} (P{\scriptsize Y}PDR is not designed to
operate at these temperatures).

In each cell there are four parameters that are important for setting
the chemical conditions and thus the (non-equilibrium) heating and
cooling rates: gas number density of H nuclei ($n_\mathrm{H}$), gas
temperature ($T$), local FUV intensity ($G_0$) and CR ionization rate
($\zeta$). The local FUV intensity is derived given $G_0$ (which may
vary with galactocentric radius) and the visual extinction, $A_V$, to
a cell. The method of \citet{W15} involves approximating $A_V$ as a
local, monotonically rising function of density $n_\HH$. The
reliability of this approximation on the scales of GMCs has been
investigated by comparison with full radiative transfer simulations of
Bisbas et al. (in prep.)  and found to be accurate to $\sim$20\% for
heating and cooling rates at typical densities of $n_{\rm
  H}=10^3\:{\rm cm^{-3}}$ and temperatures of $T=100\:$K. More
generally, on galactic scales, the validity of local approximations to
account for FUV radiative shielding in multi-dimensional simulations
has been studied comparing to a more accurate ray-tracing-based
approach \citep{SKKO17} and is reported to be physically appropriate
to model the thermal impact of the HI--${\rm H_2}$ and CII/CI/CO
transitions.

In this paper we have extended the grid of heating and cooling
functions to span extra dimensions of $G_0$ and $\zeta$, i.e., up to a
four dimensional parameter space. The range of background diffuse FUV
intensities and cosmic ray ionization rates are set as described in
\S\ref{sec:nc}. The heating/cooling functions span the density and
temperature space of $10^{-4}< n_\mathrm{H}/\mathrm{cm}^{−3}
< 10^6$ and $2.7< T/\mathrm{K}< 10^7$,
covering our regime of interest.

The resulting heating/cooling rates and corresponding mean molecular
weights are incorporated into \enzo\ via the \textit{Grackle} external
chemistry and cooling library \citep{Smith17}. The information is read
in via the purely tabulated mode using quadrilinear interpolation to
fill the entire 4-D parameter space, and modifies the specific gas
internal energy $e=p/(\gamma -1)$ of a given cell with a net
heating/cooling rate calculated by
\begin{equation}
H = n_\HH[\Gamma-n_\HH\Lambda] \mathrm{\ erg\ cm^{-3}\ s^{-1}},
\end{equation}
where $\Gamma$ is the heating rate and $\Lambda$ is the cooling rate.
The species abundances and their line emissivities are also used to
make predictions of emission line diagnostics (discussed in the next
subsection).

\subsection{Observational Diagnostics}
\label{sec:obm}

One important output of the PDR model described in \S\ref{sec:tp} is
the detailed information about specific components that contribute to
the radiative cooling rates. Given the local volume emissivities, $j$,
of particular lines such as [CII] emission at 158~$\rm \mu m$ and
rotational lines of $^{12}\mathrm{CO}$ and $^{13}\mathrm{CO}$, we are
able to create synthetic integrated intensity maps in post-processing.

We note that radiative transfer is incorporated, approximately, in
each cell through implementation in the 1D PDR models, while (for the
results presented in this paper) full line radiative transfer is not
calculated during post-processing, i.e., we sum contributions along
lines of sight, which is only fully valid in the optically thin
limit.

Thus we typically choose to study lines in which optical 
depths should be relatively small. The resulting intensities are
simply integrated directly through the simulation domain. We also note
that CO freeze-out onto dust grains is not yet treated in our PDR
models. Following equation (6) in \citet{W15}, the integrated intensity
is calculated through
\begin{equation}
\int T_\mathrm{mb} \diff v = \frac{\lambda ^3}{2k_B} I = \frac{\lambda ^3}{2\pi k_B}\int_{\mathrm{LOS}} j \diff z,
\end{equation}
where $T_\mathrm{mb}$ is the main-beam temperature, $j$ is the emissivity (the output of PDR codes is $4\pi j$), $v$ is the
line-of-sight velocity, $\diff z$ is the line element along the line
of sight, and $\lambda$ is the chosen line centroid wavelength.

\subsection{Cloud Tracking}

The clouds in the galactic disk were located using a number density of
H nuclei threshold of $n_{\rm \HH,GMC} = 100\ \mathrm{cm^{-3}}$, i.e.,
the same value used in the TT09 study and similar to the
volume-averaged mean densities of typical Galactic GMCs. We will also
justify this choice of density threshold based on the PDR models
implemented in our simulations. When we refer to ``clouds'' or
``GMCs'' we are describing the gas that has achieved densities of at
least $n_{\rm \HH,GMC}$.

Clouds are identified utilizing \textit{k-d tree}, a
binary-space-partitioning data structure especially effective in the
proximity/clustering search, through the following loops, which
achieve the same results as the method developed by TT09:
\begin{itemize}
\item \textit{Loop 1}: local density peaks are identified within the
  cells with $n_{\HH}> n_{\rm \HH,GMC}$. If the distance
  between two peaks is smaller than the deblending length (here we set
  it to a fixed distance of 32 pc, including in the run with 4~pc
  resolution), then only the peak with the higher density is retained
  as a ``peak.''
\item \textit{Loop 2}: for every peak, adjacent cells with $n_{\HH}
 > n_{\rm \HH,GMC}$ are marked as ``belonging to the peak,''
  and cells adjacent to these marked cells are recursively marked in
  the same way.  Eventually, the cells that have continuous spatial
  connections with multiple peaks are assigned to the peak closest to
  them, i.e., each cell is only assigned to one peak. With this
  method, multiple clouds can exist in the same continuous density
  structure, if it contains more than one well-separated peak.
\end{itemize}

Once clouds are identified for each simulation output, we then
calculate their physical properties and also track cloud
evolution. The clouds at $t_0$ and those at $t_1$ (typically $t_1-t_0
=1$ Myr) are linked as parent--child pairs through the following
steps:
\begin{itemize}
\item \textit{Step 1}: assuming constant velocity within 1 Myr, a
  predicted center of mass  at $t_1$ for each cloud found at 
  $t_0$ is calculated.
\item \textit{Step 2}: a volume of radius 50 pc, larger than the
  expected deviations due to typical accelerations, centered on the
  predicted position of each cloud is searched for clouds present at
  time $t_1$. If multiple clouds are found in this region at $t_1$,
  the nearest one is linked with the cloud at $t_0$.
\item \textit{Step 3}: in the case that multiple $t_0$ clouds
  (parents) are linked to the same $t_1$ cloud (child) based on the
  previous step, the child will be assigned to the ``parents'' with
  the nearest predicted $t_1$ location. Other ``parents'' will repeat
  step 2 and search for alternative children excluding the assigned
  one until no conflict exists, or no alternative child exists so that
  the children defined in step 2 have to be linked to those
  ``parents.''
\item \textit{Step 4}: if no clouds at $t_1$ are linked with the $t_0$
  cloud, then a volume with radius equal to $3\times$ the average
  radius (defined as $R_c=\sqrt{A_c/\pi}$ where $A_c$ is projected
  area of the cloud in the $x$-$y$ plane) of the $t_0$ cloud is
  searched. This allows for large, extended clouds whose centers may
  have shifted due to external perturbations.
\item \textit{Step 5}: if multiple clouds at $t_0$ are linked to one
  cloud at $t_1$, then the most massive $t_0$ cloud is claimed to
  survive while others are ``destroyed'' by merger. If one $t_0$ cloud
  has no child, then it is claimed to have been destroyed by
  non-merger processes. Cloud formation is claimed for each $t_1$
  cloud without any parent.
\end{itemize}

This procedure tracks merger events accurately, but we caution that
this simple prescription for identifying destruction events may have
difficulties, particularly when strong feedback mechanisms
are eventually included. Still, the prescription works well in the
current simulation, since we have limited cloud destruction, i.e., the
diffuse FUV radiation does not lead to significant GMC destruction.

\section{Large-Scale ISM Properties}
\label{sec:ism}

In the following sections, we restrict our analysis of ISM and GMC
properties to galactocentric radii between 2.5 and 8.5 kpc in order to
avoid structures affected by the boundaries of the main disk.

\subsection{Global Disk}

\begin{figure*}[htb!]
\centering
\includegraphics[width= 1.0 \textwidth]{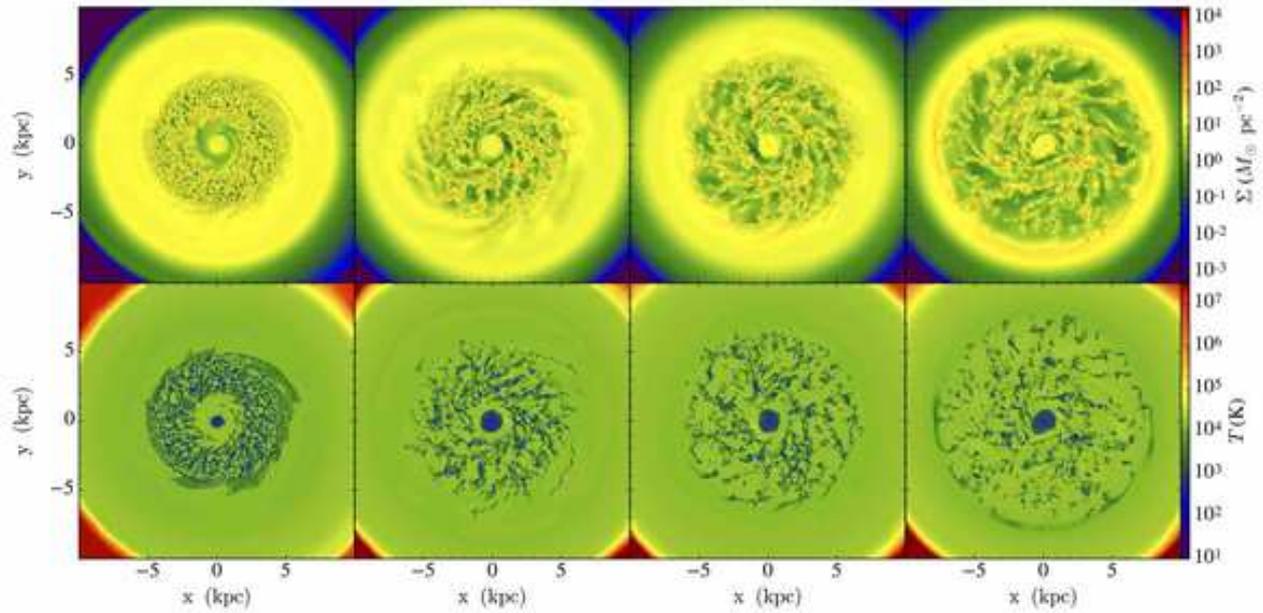}
\caption{
Evolution of the galactic disk for Run I ($G_0 = 4$) (100, 200, 250
and 300 Myr from left to right). Images are 20 kpc across and show the
disk gas mass surface density $\Sigma$ (top row) and mass-weighted
temperature $T$ (bottom row), integrated vertically over the disk
(over a range $-1\:{\rm kpc}<z<+1\:{\rm kpc}$).}
\label{fig1}
\end{figure*}

\begin{figure*}[htb!]
\centering
\includegraphics[width= 1.0 \textwidth]{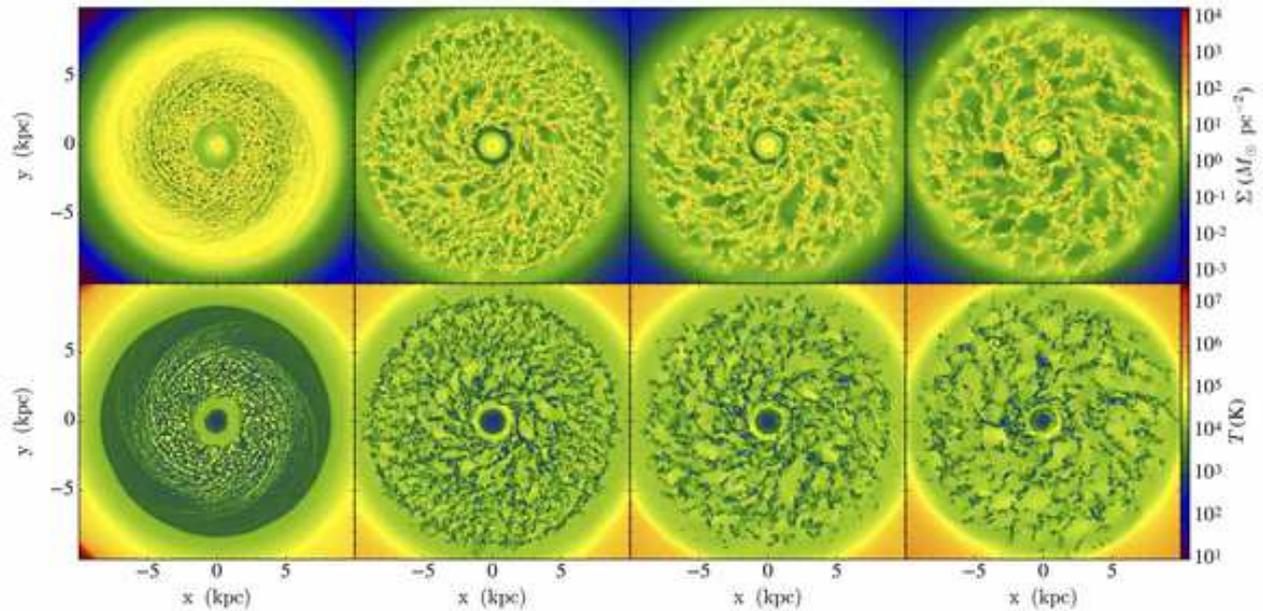}
\caption{Same as Fig.~\ref{fig1}, but for Run II ($G_0 = 1$).}
\label{fig2}
\end{figure*}

\begin{figure*}[htb!]
\centering
\includegraphics[width=1.0 \textwidth]{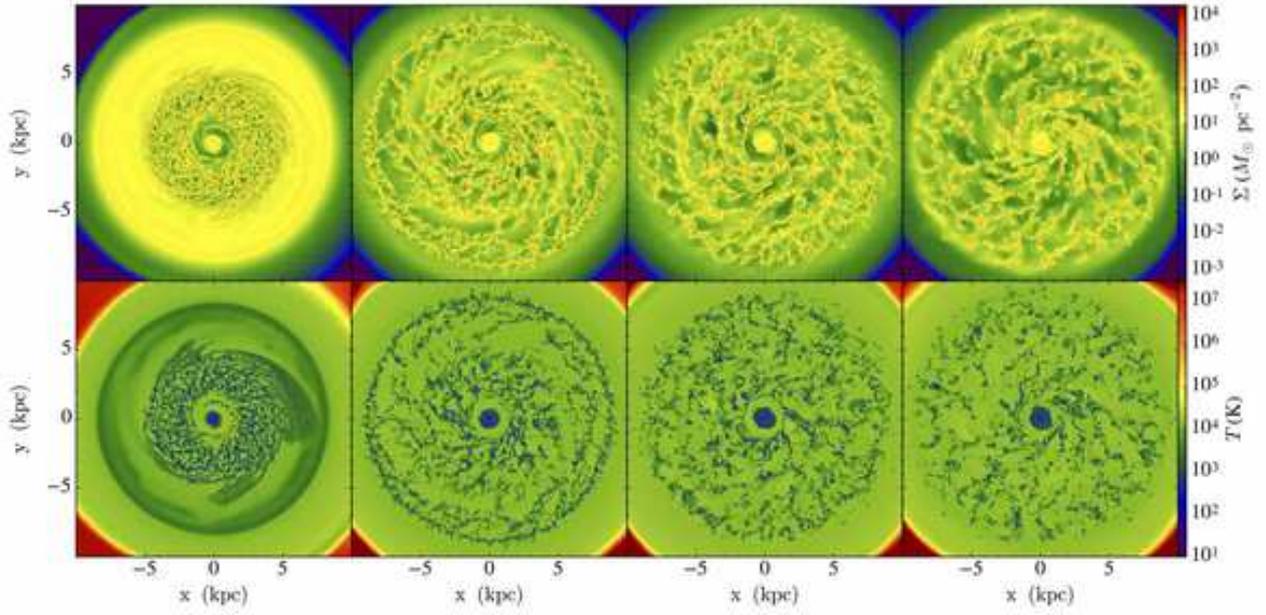}
\caption{Same as Fig.~\ref{fig1}, but for Run III (radially variable $G_0$).}
\label{fig3}
\end{figure*}

\begin{figure*}[htb!]
\centering
\includegraphics[width= 1.0 \textwidth]{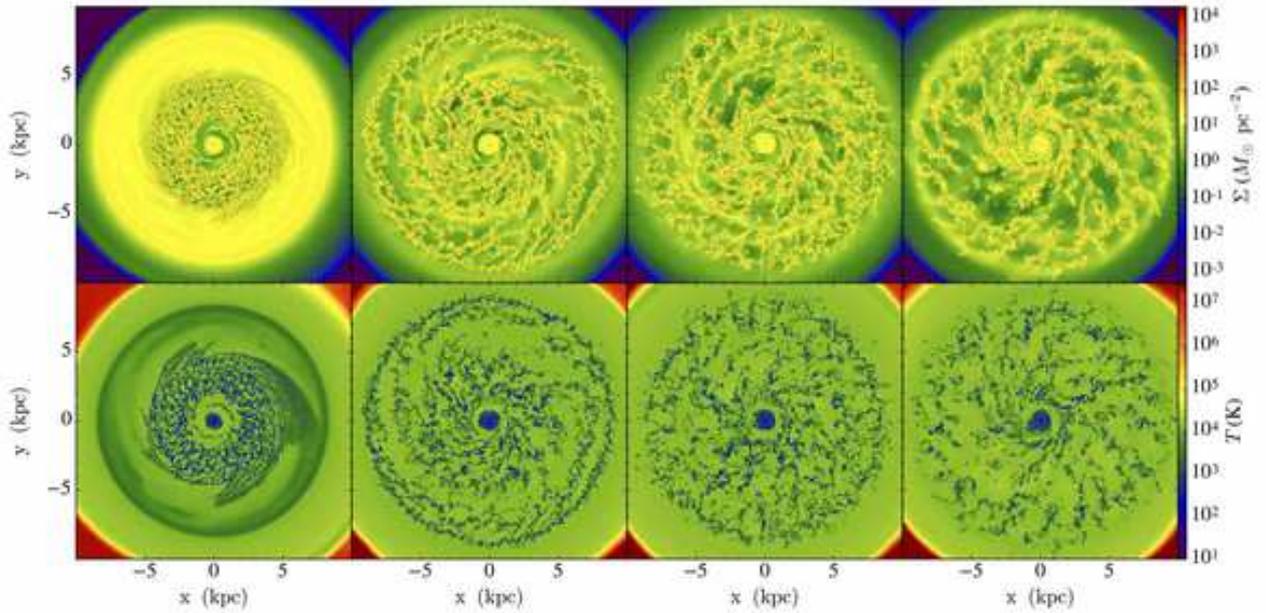}
\caption{Same as Fig.~\ref{fig1}, but for Run IV (radially variable $G_0$ with 10~K temperature floor).}
\label{fig1-2}
\end{figure*}

\begin{figure*}[htb!]
\centering
\includegraphics[width= 1.0 \textwidth]{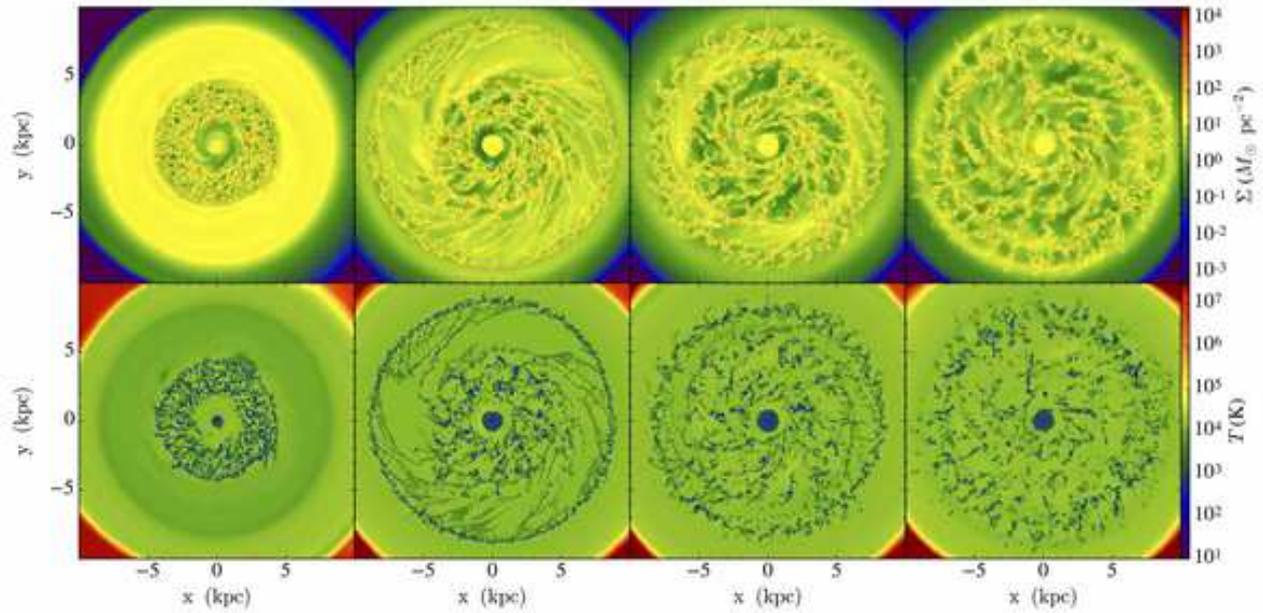}
\caption{Same as Fig.~\ref{fig1}, but for Run V (radially variable $G_0$ and $\zeta$ with 10~K temperature floor).}
\label{fig2-2}
\end{figure*}

\begin{figure*}[htb!]
\centering
\includegraphics[width=1.0 \textwidth]{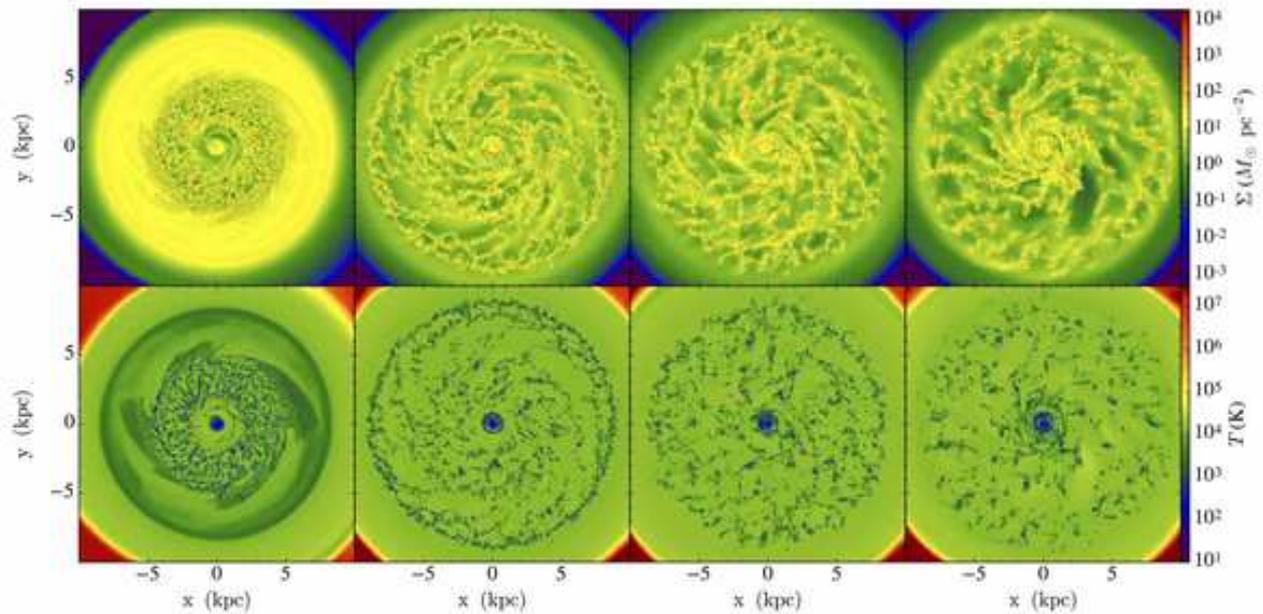}
\caption{Same as Fig.~\ref{fig1}, but for Run VI (radially variable $G_0$ with 10~K temperature floor and 4~pc resolution).}
\label{fig3-2}
\end{figure*}

The time evolution of the galactic disks of Runs I to VI are shown in
Figures \ref{fig1}, \ref{fig2}, \ref{fig3}, \ref{fig1-2}, \ref{fig2-2}
and \ref{fig3-2}, respectively. At the beginning, with an initialized
velocity dispersion of $\sigma_g = c_s \simeq 8\ \mathrm{km\ s^{-1}}$, 
the disks are marginally stable. As a result of radiative cooling, the
disks become unstable and fragment into overdense structures, which
increases the self-shielding from the diffuse FUV radiation, raises
line cooling rates, thus leading to more rapid cooling and gathering
of material from the surroundings. As overdense regions become denser
than $n_{\rm \HH,GMC} = 100\: \mathrm{cm^{-3}}$, they are identified
as components of GMCs. Similar to the results seen by TT09, this
evolution, driven by cooling, occurs most rapidly in the denser inner
regions and therefore the first structure to form is an overdense ring
near the inner boundary due to the Toomre ring instability. The
instability gathers material from a range of radii about equal to the
Toomre length, $\lambda_T=2\pi^2 G\Sigma_g/\kappa^2$. The instability
then propagates to outer regions as radiative cooling
continues. Overdense local spiral patterns form and fragment into
clouds.

Since the early phase evolution is quite artificial, in Figures
\ref{fig1} to \ref{fig3-2} we focus on later stages from 100 to
300~Myr, showing also intermediate stages at 200 and 250~Myr. Later we
will be focussing on the statistics of ISM and GMC properties from 200
to 300~Myr.

Comparing the first three runs (Figures \ref{fig1} to \ref{fig3}),
which have different FUV intensities, we see that energy injection
from FUV radiation suppresses the growth of gravitational
instabilities, especially in the lower density outer regions, thus
shaping the spatial distribution as well as the global evolutionary
history of GMCs. On the global scales viewed in Figures \ref{fig1} to
\ref{fig3-2}, there appear to be only modest effects on the dense gas
structures caused by lowering the temperature floor from 300~K to
10~K, i.e., comparing Run III and Run IV. Applying a radially variable
$\zeta$ in Run V, which overall has stronger heating than the constant
$\zeta$ case, we see that fragmentation is delayed to some extent. CR
heating is of greater importance in the denser, molecular phase where
FUV radiation is shielded and equilibrium temperatures are lower.

\begin{figure*}[htb!]
\centering
\includegraphics[width = 1.0 \textwidth]{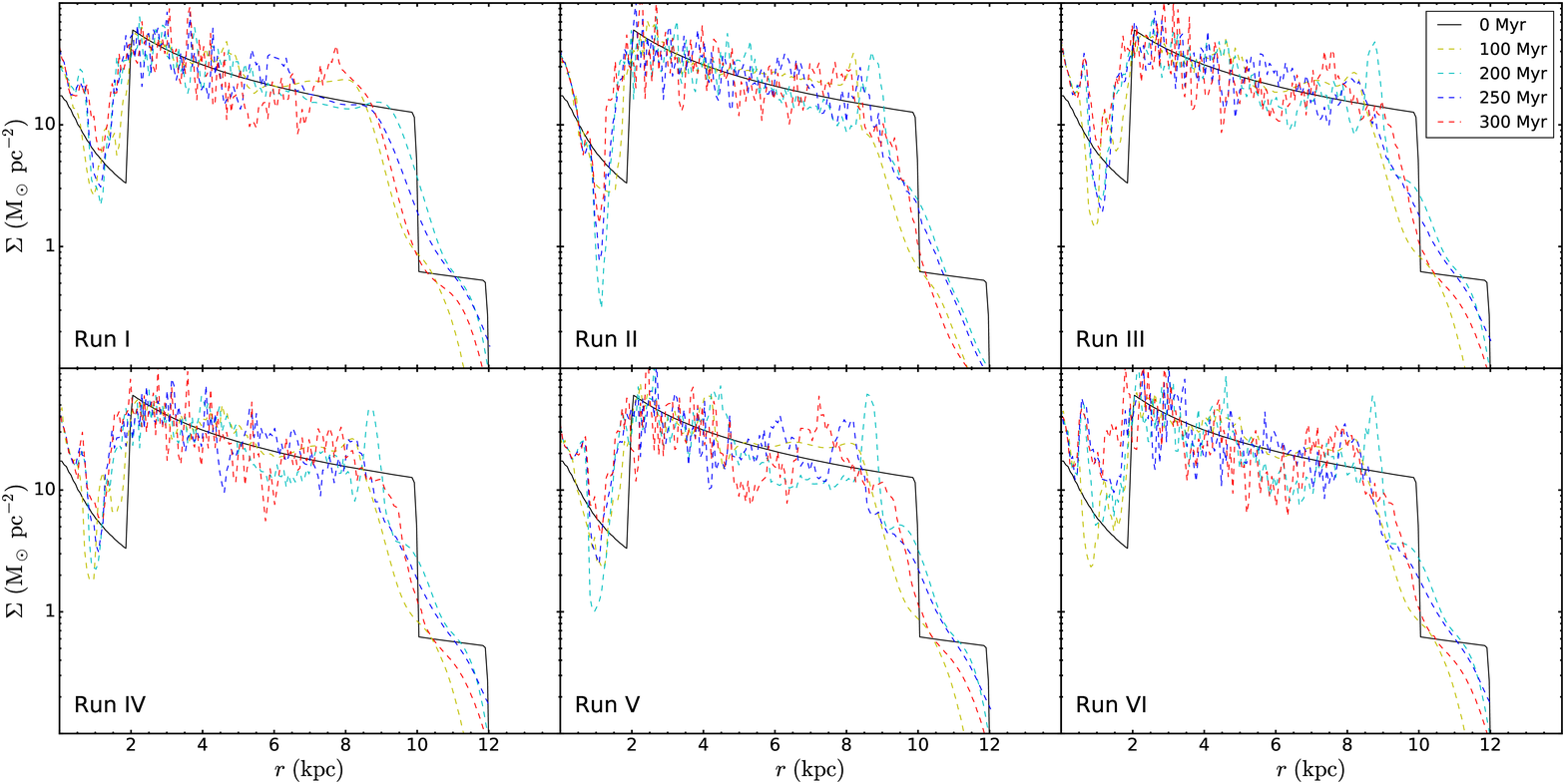}
\caption{
The evolution of azimuthally averaged radial profiles of gas mass surface density
$\Sigma=\int_{-1\ \mathrm{kpc}}^{+1\ \mathrm{kpc}}\rho dz$ for Run I
(\textit{upper left}, $G_0$=4), II (\textit{upper middle}, $G_0$=1),
III (\textit{upper right}, radially variable $G_0$), IV (\textit{lower
  left}, radially variable $G_0$, 10~K temperature floor), V
(\textit{lower middle}, radially variable $G_0$ and $\zeta$, 10 K
temperature floor), VI (\textit{lower right}, radially variable $G_0$,
constant $\zeta$, 10 K temperature floor with 4 pc resolution).}
\label{fig:Sigma_r}
\end{figure*}

Figure~\ref{fig:Sigma_r} shows the time evolution of the radial
profiles of $\Sigma$, azimuthally-averaged in annuli, in these
simulations. Fluctuations on the order of factors of a few are seen to
develop due to gravitational instabilities. However, overall the
average values remain similar to those of the initial conditions, with
the exceptions being caused by modest amounts of radial accretion seen
most obviously at the boundaries of the main disk region. These
results are relatively insensitive to the choices of PDR-related
parameters. 

\begin{figure*}[htb!]
\centering
\includegraphics[width = 1.0 \textwidth]{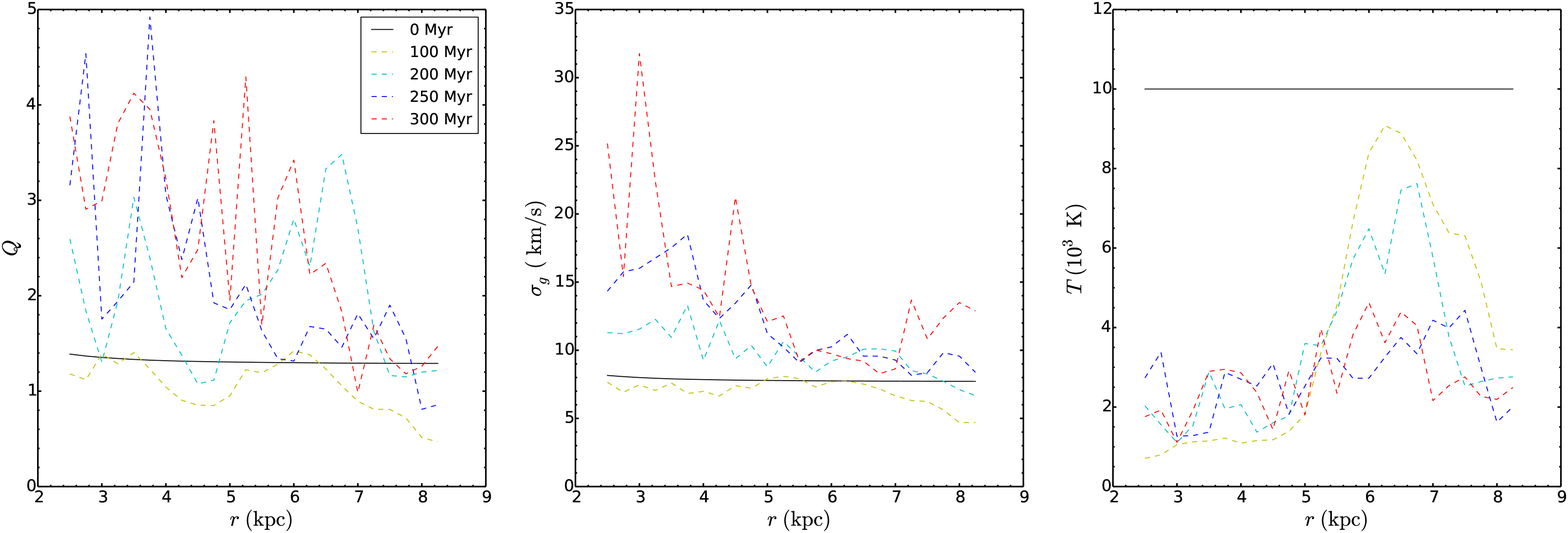}
\caption{
Galactic disk azimuthally averaged radial profiles (2.5 kpc$<r<$ 8.5
kpc) and their evolution for Run VI; from left to right: (a) Toomre
$Q$ parameter, (b) one-dimensional gas velocity dispersion, $\sigma_g$
(mass-weighted average over -1 kpc$<z<$ 1 kpc), (c) gas temperature
(mass-weighted average over -1 kpc$<z<$ 1 kpc).}
\label{fig:Q_r}
\end{figure*}

The time evolution of the azimuthally-averaged radial profiles (in
range 2.5 kpc$<r<$ 8.5 kpc, with annuli of width 0.5~kpc) 
of the Toomre $Q$ parameter, the 1D gas velocity dispersion
$\sigma_g$, and average gas temperature $T$ for Run VI are shown in
Figure~\ref{fig:Q_r}. Note that $\sigma_g=\sqrt{c_s^2+\sigma^2_{\rm
    nt}}$ is a mass-weighted average over -1 kpc$<z<$ 1 kpc, where
$\sigma_{\rm nt}$ is the 1D velocity dispersion of the gas motions in
the plane of the disk after the subtraction of the local circular
velocity, $T$ is a mass-weighted average over -1 kpc$<z<$ 1 kpc and
$Q$ is calculated via
\begin{equation}\label{eq:Q}
Q = \frac{\kappa \sigma_g}{\pi G \Sigma}.
\end{equation}
During the early stage of the simulation, the gas cools rapidly,
causing $Q$ to drop and leading to gravitational instability. The
inner region, where gas is denser,
has faster cooling, so GMCs form earlier. We will see below that disk
fragmentation eventually (by $\sim 200\:$Myr) reaches a relatively
steady state, with the ISM existing in two main phases. The diffuse
gas, being more exposed to the FUV field, retains a relatively high
temperature and so the overall mass-weighted temperature is $\gtrsim
2,000\:$K, even though the majority ($\sim2/3$) of the mass is in
dense, ``GMC''-like structures. However, $\sigma_g$ is dominated by
the random motions of GMCs in the disk, i.e., $\sigma_{\rm nt}$. These
are accelerated by mutual gravitational interactions between
self-gravitating GMCs and damped by cloud-cloud collisions (Gammie et
al. 1991). Thus at late times, $Q$, as defined by equation
(\ref{eq:Q}), rises to values of $\sim3$, i.e., it is already in a
highly fragmented state due to gravitational instability, but then has
the superficial appearance of being gravitationally stable.

\begin{figure}[htb!]
\centering
\includegraphics[width = 0.475 \textwidth]{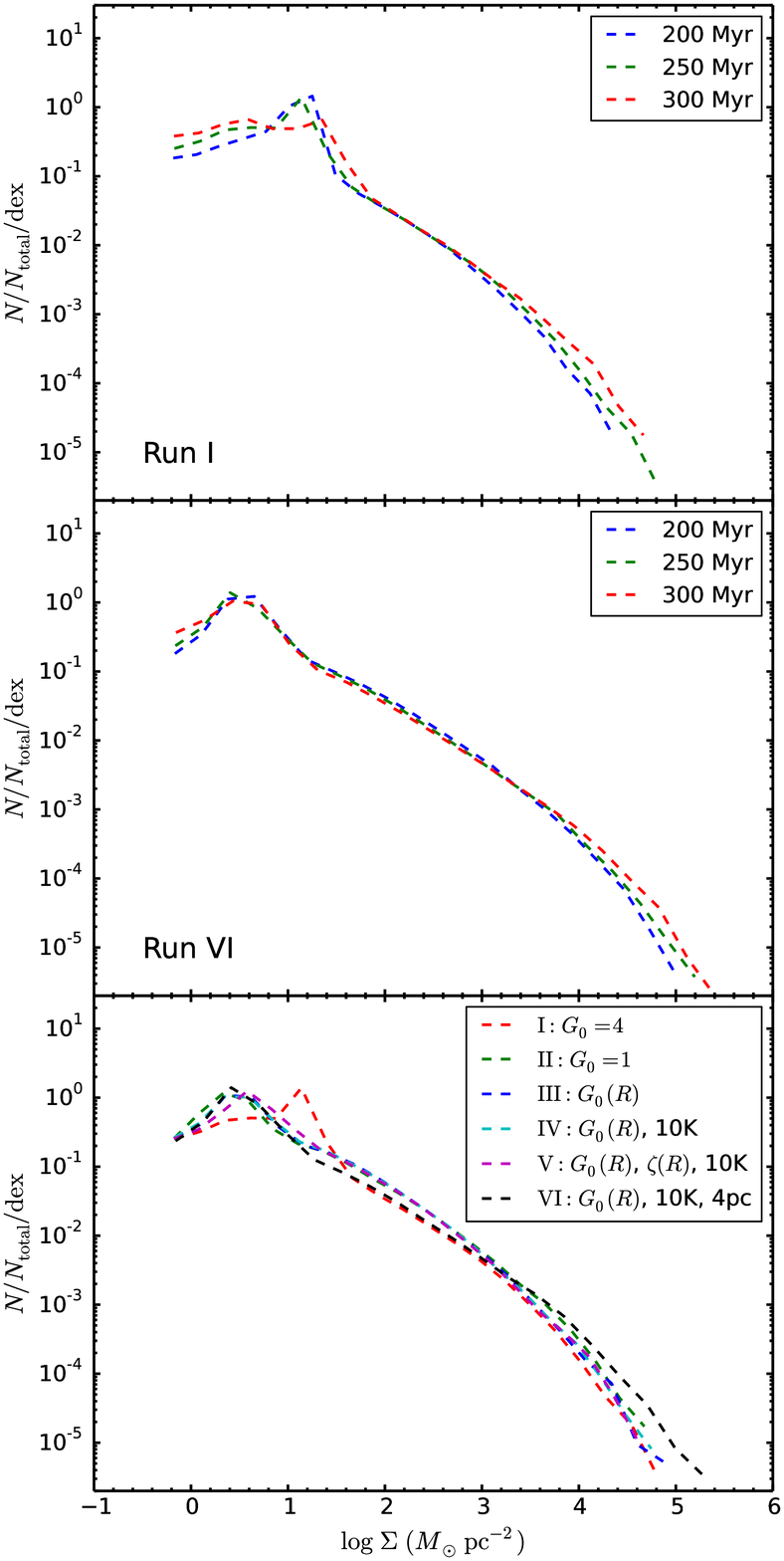}
\caption{
PDFs of gas mass surface density, $\Sigma$, as viewed from above the
disk, evaluated from 2.5 kpc $<\ r\ <$ 8.5 kpc and projecting through
-1 kpc $<\ z\ <$ 1 kpc.
\textit{Top panel}: results for Run I ($G_0=4$) at 200 (blue), 250
(green) and 300 (red) Myr.
\textit{Middle panel}: results for Run VI (Variable $G_0$) at 200
(blue), 250 (green) and 300 (red) Myr.
\textit{Bottom panel}: results for Runs I (red), II (green), III
(blue), IV (cyan), V (magenta) and VI (black) at 250~Myr.}
\label{fig:SigmaPDF}
\end{figure}

The top panel of Figure \ref{fig:SigmaPDF} shows the evolution of the
$\Sigma$ PDF of Run I from 200 to 300 Myr. Overall this PDF stays
relatively constant during this period, implying that the ISM of the
global disk is approaching a quasi statistical equilibrium. However,
there is a tendency seen for the mass fraction in the densest regions
to be increasing. The effect of the final spreading of fragmentation
to the outer disk by 300~Myr is also apparent in reducing the peak
near $\Sigma=10\:M_\odot\:{\rm pc}^{-2}$. The middle panel shows
equivalent results for Run VI, which is closer to full fragmentation
during the period from 200 to 300~Myr. These PDFs show a greater
degree of constancy, with just a minor enhancement in the fraction of
material at the highest mass surface densities as the disk
evolves. However, examining the morphologies shown in
Fig.~\ref{fig3-2} we do see a trend for apparent agglomeration of
clouds into larger structures. This will be examined further in terms
of GMC mass functions in the next section. Finally, the bottom panel
of Figure \ref{fig:SigmaPDF} compares the $\Sigma$ PDFs of all six
runs at 250~Myr. Overall the results are quite similar between the
different runs, with the case of $G_0=4$ having the largest
differences, as fragmentation in the outermost regions is still
ongoing. The result of Run VI, which has the highest resolution, also
shows a difference in being able to resolve a higher fraction of mass
at the greatest mass surface densities.

\begin{figure*}[htb!]
\centering
\includegraphics[width = 1.0 \textwidth]{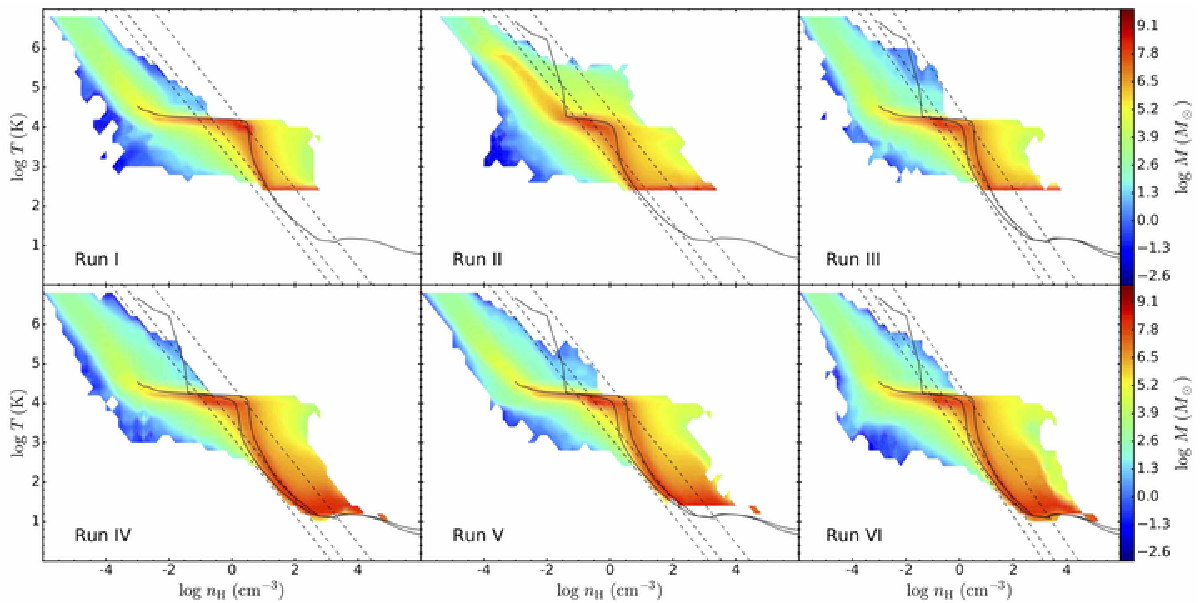}
\caption{
Density versus temperature phase plots showing the distribution of gas
mass at 300 Myr for Run I ($G_0=4$), II ($G_0=1$), III (radially
variable $G_0$), IV (radially variable $G_0$ with 10~K temperature
floor), V (radially variable $G_0$ and $\zeta$ with 10~K temperature
floor), VI (radially variable $G_0$ with 10~K temperature floor and
4~pc resolution). The mass scale indicates the total mass of gas
within a given $\Delta {\rm log}\:n_{\rm H}$ by $\Delta {\rm log}\:T$
interval.
In each panel, the solid curves show the equilibrium temperature as a
function of density.  For Runs III to VI, the upper and lower solid
curves at $\mathrm{n_H} > 10^{-1.5}$ $\mathrm{cm^{-3}}$ represent the
equilibrium temperature for $G_0=4$ and $G_0=1$, respectively, both
with $\zeta=10^{-16}\:{\rm s}^{-1}$.
The dashed lines show estimates of the total pressure in the Milky Way
(Boulares \& Cox 1990), $P_{\mathrm{tot}}/k = 2.8\times10^4
\mathrm{\ K\ cm^{-3}}$ (top line), the total thermal pressure,
$P_{\mathrm{th}}/k = 0.36\times10^4 \mathrm{\ K\ cm^{-3}}$ (middle
line), and the thermal pressure excluding the hot gas component,
$P_{\mathrm{th,nohot}}/k = 0.14\times10^4 \mathrm{\ K\ cm^{-3}}$
(bottom line).}
\label{fig:rhoT}
\end{figure*}

The density-temperature phase plots of the ISM at 300 Myr are shown in
Figure \ref{fig:rhoT} for all runs. The solid curves delineate the
equilibrium temperature as a function of density, set by heating and
cooling functions described in \S\ref{sec:method}. Much of the mass
(especially the densest parts) of the simulated ISM is at a
temperature close to that expected from thermal equilibrium. Some
deviations are expected due to energy injection and dissipation via
dynamical processes such as heating due to compressive motions, e.g.,
cloud-cloud collisions, and cooling due to rarefactions.

Comparing Run IV to Run V, which has a radially variable and overall
stronger $\zeta$, we can see that CR ionization is an important
heating source at low temperatures ($\lesssim30$ K), which affects the
lowest temperatures that the dense gas can reach. Additionally,
comparing Run IV and Run VI, one notices in the latter with its higher
resolution a somewhat larger fraction of dense gas is seen to deviate
from thermal equilibrium due to dynamical effects.

The typical local Milky Way total diffuse ISM pressure is about
$2.8\times 10^4\:$K $\mathrm{cm^{-3}}$ and its thermal components are
about an order of magnitude smaller \citep{BoularesCox90}. These
pressures are shown by straight lines in Figure
\ref{fig:rhoT}. Incorporating diffuse FUV feedback, our simulated
diffuse ISM has comparable thermal pressures to the observed Milky Way
thermal pressures at densities $\gtrsim 10^{-1}\ \mathrm{cm^{-3}}$. At
lower densities the simulation pressures are lower, which is most
likely due to the lack of supernova feedback that would create a hot
ionized medium with $T\gtrsim 10^6\:$K. At high densities, GMCs in the
simulation are at much higher pressures than the diffuse ISM, mainly
due to their self-gravity.

However, recall that in addition to the lack of SN feedback, the
simulations also lack localized FUV, EUV and wind momentum feedback
from young stars. They also lack magnetic fields and the dynamical
effects due to cosmic ray pressure gradients. Both of these are
important pressure components in the Milky Way. The inclusion of these
physical processes is deferred to future papers in this series. Thus
when interpreting the results presented here, deviations from
simulated and observed ISM properties may be due to the lack of these
physical processes.

\begin{figure}[htb!]
\centering
\includegraphics[width=0.5 \textwidth]{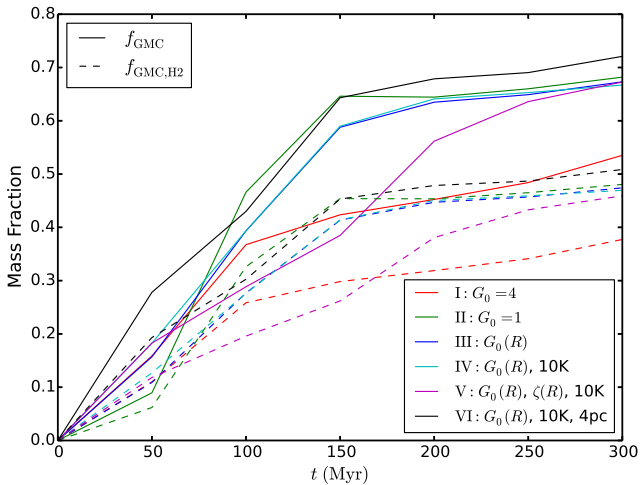}
\caption{
Fraction of total disk mass in ``GMCs,'' i.e., with $n_{\rm
  H}>100\:{\rm cm^{-3}}$ ($f_{\rm GMC}$) as a function of simulation
time (solid lines) for Run I (red), II (green), III (blue), IV (cyan),
V (magenta) and VI (black). The dashed lines show the fraction of the
total disk mass that is in ``GMCs'' and is in the $\rm H_2$ phase
($f_{\rm GMC,H2}$).}
\label{fig:gmcmassfrac_t}
\end{figure}

We evaluate gas density distributions and compositions for the main
region of interest of the disk (2.5 kpc $<\ r\ <$ 8.5 kpc and -1 kpc
$<\ z\ <$ 1 kpc).  The mass fraction of gas that is in GMCs (i.e.,
with $n_{\rm H}\geq 100\:{\rm cm^{-3}}$), $f_{\rm GMC}$, is shown as a
function of time in Figure~\ref{fig:gmcmassfrac_t} (solid
lines). These fractions rise for the first $\sim150$~Myr and then
approximately stabilize. The case with $G_0=4$ has the smallest
fraction of gas in GMCs, about 45\% from 200 to 300~Myr. The other
models have higher fractions, which are also very similar to each
other, i.e., with about 67\% of gas mass in GMCs. We note that this
fraction is also very similar to that found by TT09, even though this
simulation did not include FUV feedback. This tells us that the
formation of dense gas, driven by gravitational instability, is not
that sensitive to FUV feedback, until $G_0\sim4$, when the outer disk
begins to be stabilized.

Figure~\ref{fig:gmcmassfrac_t} also shows the fraction of the total
gas in the disk that is in GMCs and also in the $\rm H_2$ phase,
$f_{\rm GMC,H2}$ (dashed lines). This is the result of our PDR
modeling of the gas with densities $n_{\rm H}\geq100\:{\rm
  cm^{-3}}$. This fraction rises with a similar time dependence as
$f_{\rm GMC}$. The ratio of the dashed and solid lines for a given
simulation, i.e., $f_{\rm GMC,H2}/f_{\rm GMC}$, is the average $\rm
H_2$ mass fraction within our defined ``GMCs.'' We see that this ratio
is close to 70\%. This is broadly consistent with the fact that
Galactic GMCs are known to have large mass fractions in atomic
envelopes (\cite{Blitz90,Wannier91}). Thus our choice of $n_{\rm
  H}= 100\:{\rm cm^{-3}}$ as a threshold density to define ``GMCs''
appears reasonable.

\begin{figure}[htb!]
\centering
\includegraphics[width=0.5 \textwidth]{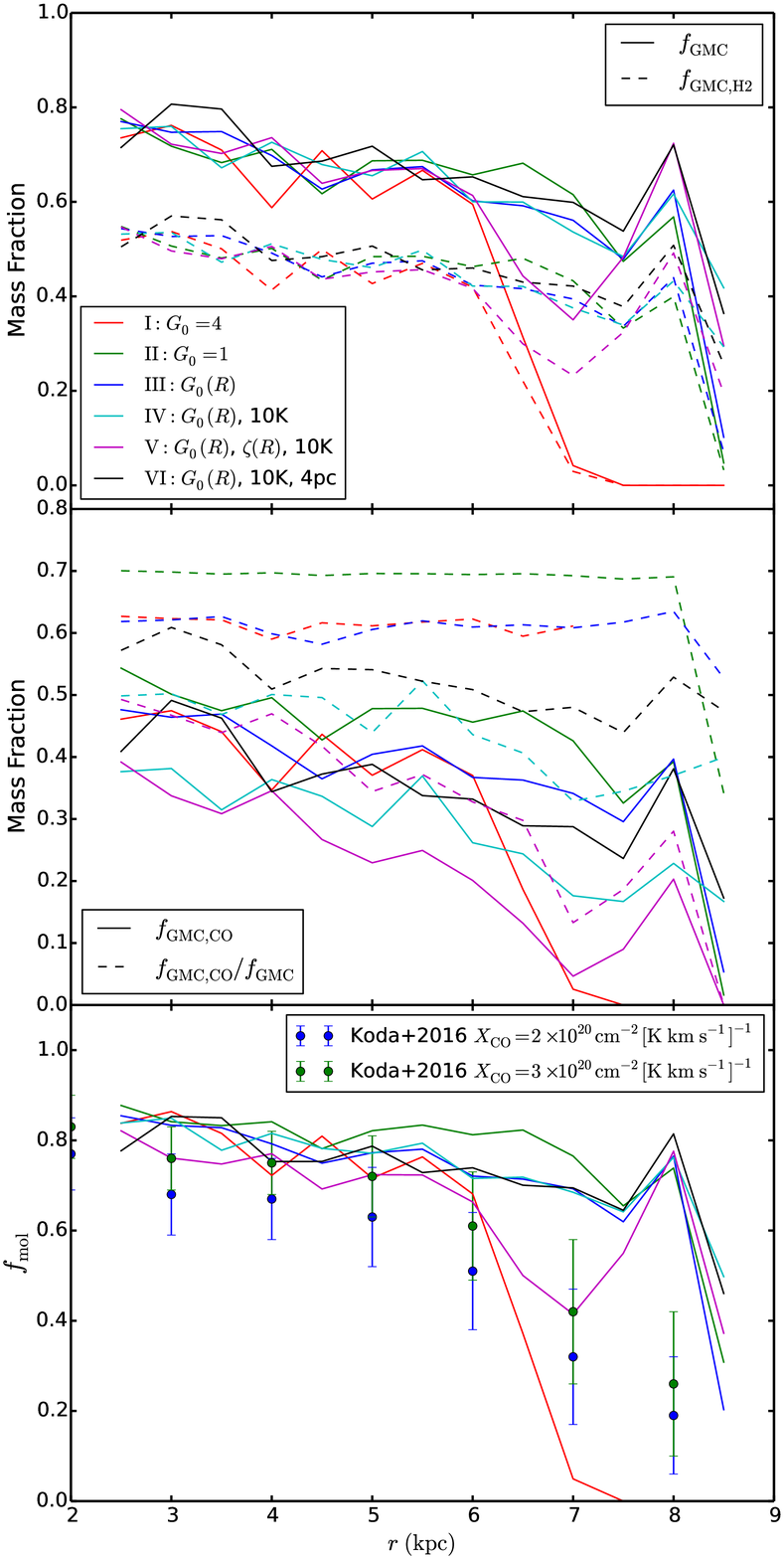}
\caption{
{\it (a) Top panel:} Fraction of total disk gas mass in ``GMCs''
($f_{\rm GMC}$), i.e., with $n_{\rm H}>100\:{\rm cm^{-3}}$, as a
function of galactocentric radius (solid lines) for Runs I to VI as
indicated in legend, all evaluated at 250~Myr. The dashed lines show
the fraction of the total disk gas mass in the ``GMCs'' that is also
in the $\rm H_2$ phase ($f_{\rm GMC,H2}$).
{\it (b) Middle panel:} Fraction of total disk gas mass in GMCs that
is also ``CO-rich,'' i.e., with $n_{\rm CO}/n_{\rm H}>10^{-5}$
($f_{\rm GMC,CO}$ - solid lines), with results shown for the runs at
250~Myr (see legend in panel a). The ratio $f_{\rm GMC,CO}/f_{\rm
  GMC}$ is also shown (dashed lines).
{\it (c) Bottom panel:} Molecular fraction, $f_{\rm mol}\equiv
\Sigma_{\rm H2}/(\Sigma_{\rm H2}+\Sigma_{\rm HI})$ for the different
runs (see legend in panel a), compared to the Milky Way from the
analysis of Koda et al. (2016) for two different values of $X_{\rm
  CO}$, as indicated.
}
\label{fig:gmcmassfrac_r}
\end{figure}

Figure \ref{fig:gmcmassfrac_r}a shows the radial profiles of $f_{\rm
  GMC}$ and $f_{\rm GMC,H2}$ at 250~Myr. These fractions decline
gradually as one goes from the denser, inner regions to the lower
density, outer regions of the disk. In Run I, i.e., with high $G_0$,
the transition from a gravitationally unstable inner
($\lesssim6.5\:$kpc) zone to a stable outer region is clearly seen in
these profiles. The other simulation runs are seen to have radial
profiles of these metrics that are quite similar to each other.

In Figure \ref{fig:gmcmassfrac_r}b with solid lines, we show the
galactic radial profiles at 250~Myr of the fraction, $f_{\rm GMC,CO}$,
of the total disk mass that is in ``GMCs'' and has a CO abundance
$n_{\rm CO}/n_{\rm H}>10^{-5}$, i.e., within an order of magnitude of
the maximal CO abundance of $n_{\rm CO}/n_{\rm H}=10^{-4}$, given our
adopted total C and O abundances (i.e., $n_{\rm C} / n_{\rm H}
=1\times 10^{-4}$). Like $f_{\rm GMC}$ and $f_{\rm GMC,H2}$, the
profiles of $f_{\rm GMC,CO}$ show a general decrease with $r$, but now
starting from values of $\sim0.5$ in the inner disk and reaching
$\sim0.1$ (or less in the case of Run I) in the outer disk. As
expected, Run II with a low value of $G_0=1$ has the highest values of
$f_{\rm GMC,CO}$. Run V, with overall strongest CR ionization rate,
generally shows the lowest $f_{\rm GMC,CO}$, which we attribute to the
fact that CO is efficiently destroyed indirectly by CRs via creation
of He$^+$ \citep{Bisbas15}.
Figure~\ref{fig:gmcmassfrac_r}b also shows the ratio $f_{\rm
  GMC,CO}/f_{\rm GMC}$ (dashed lines), which represents how well CO
traces the defined ``GMCs.'' 
Comparing the first three runs, a stronger FUV radiation leads to
slightly lower values of $f_{\rm GMC,CO}/f_{\rm GMC}$. Here we again
see that higher values of CRs in Run V compared to Run IV over most of
the disk lead to reduced fraction of the GMCs that are well traced by
CO. In general, the structures identified as ``GMCs'' can have
significant, $\sim1/3$ mass fractions in ``CO-dark'' regions (and
even more in Run V), consistent with the properties of Milky Way GMCs
(e.g., \cite{Wolfire10}).

Figure \ref{fig:gmcmassfrac_r}c shows the radial profile of $f_{\rm
  mol}\equiv \Sigma_{\rm H2}/(\Sigma_{\rm H2}+\Sigma_{\rm HI})$ in the
simulations and a comparison with its observed profile in the Milky
Way from the analysis of \citet{Koda16}. We see that there is
generally quite good agreement between the simulations and the
observational data, indicating that, at least by this metric, the
simulations are a reasonable representation of the Milky Way, and
perhaps indicating that various physical effects that are not yet
included (e.g., localized star formation feedback, including
supernovae), are not that important in regulating molecular mass
fractions.

To gauge if the ISM of the main disk region achieves a quasi
statistical equilibrium state from 200 to 300~Myr, Table~\ref{tab2}
summarizes the values of $f_{\rm GMC}$ and $f_{\rm GMC,H2}$. We see
that in most runs the ISM is in a relatively steady state during this
period, with little ($\lesssim 10\%$) change in these
metrics. However, runs I ($G_0=4$) and V (variable $G_0$ and $\zeta$)
are a little different, since the fragmentation of outer disk regions
is delayed and is continuing even after 200 Myr. Thus from 200 to
300~Myr, $f_{\rm GMC}$ increases by about 20\% in these runs.
Table~\ref{tab2} also shows the fractions of the total $\mathrm{H}_2$
mass in the main disk region that are contained in the defined GMCs,
i.e., $M_\mathrm{GMC,\mathrm{H2}}/M_\mathrm{disk,\mathrm{H2}}$. These
fractions are typically $>80\%$. We also list the fractions, $f_{\rm
  GMC,CO}$, of the disk gas mass that is in ``GMCs'' and ``CO-rich,''
and $f_{\rm GMC,CO}/f_{\rm GMC}$. Again, these quantities are
relatively stable, except for Runs I and V.


\begin{table*}[htb!]
\tbl{Properties of the ISM at 200 and 300~Myr are shown for each simulation
run. From left to right: mass fraction of total disk gas in GMCs
($f_{\rm GMC}$); mass fraction of total disk gas in GMCs and in the
$\mathrm{H_2}$ phase ($f_{\rm GMC,H2}$); $f_{\rm GMC,H2}/f_{\rm GMC}$;
GMC $\mathrm{H_2}$ phase mass compared to the whole disk's
$\mathrm{H_2}$ mass; mass fraction of total disk gas in GMCs that is
also ``CO rich,'' i.e., with $\mathrm{n_{CO}/n_{H}>10^{-5}}$ ($f_{\rm GMC,CO}$); 
mass fraction of GMC gas that is ``CO rich,'' i.e., $f_{\rm GMC,CO}/f_{\rm GMC}$.}{
\begin{tabular*}{0.9 \textwidth}{c@{\extracolsep{\fill}}|cc|cc|cc|cc|cc|cc}
\hline
Run & \multicolumn{2}{c}{$f_{\rm GMC}$} &\multicolumn{2}{c}{$f_{\rm GMC,H2}$} 
&\multicolumn{2}{c}{$f_{\rm GMC,H2}/f_{\rm GMC}$} & \multicolumn{2}{c}{$M_\mathrm{GMC,\mathrm{H2}}/M_\mathrm{disk,\mathrm{H2}}$} 
& \multicolumn{2}{c}{$f_{\rm GMC,CO}$} &\multicolumn{2}{c}{$f_{\rm GMC,CO}/f_{\rm GMC}$} \\
Time (Myr) & {200} & {300} & {200} & {300} & {200} &{300} & {200} & {300} & {200} & {300} & {200} & {300}\\
\hline
I & 0.452 & 0.535 & 0.319 & 0.377 & 0.706 & 0.705 & 0.857 & 0.859 & 0.395 & 0.468 & 0.874 & 0.875  \\
II & 0.644&  0.682 & 0.467& 0.478  & 0.706 & 0.714 & 0.791 & 0.814 & 0.634  & 0.673 & 0.984  & 0.987 \\
III & 0.634 & 0.673 & 0.447 & 0.474 & 0.705 & 0.704 & 0.844 & 0.847 & 0.547 & 0.597 & 0.863  & 0.887\\
IV & 0.641 & 0.667 & 0.450& 0.469 & 0.702 & 0.703 & 0.840 & 0.839 & 0.382 & 0.456 & 0.596 & 0.684 \\
V & 0.561 & 0.672 & 0.381 & 0.459 & 0.680 & 0.683 & 0.890 & 0.889 & 0.177 & 0.388 & 0.316 & 0.577 \\
VI & 0.678 & 0.721 & 0.479 &0.509 & 0.706 & 0.706 & 0.897 & 0.901 & 0.500 & 0.576 & 0.737 & 0.799 \\
\hline
\end{tabular*}}\label{tab2}
\end{table*}

\begin{figure*}[htb!]
\centering
\includegraphics[width = 1.0 \textwidth]{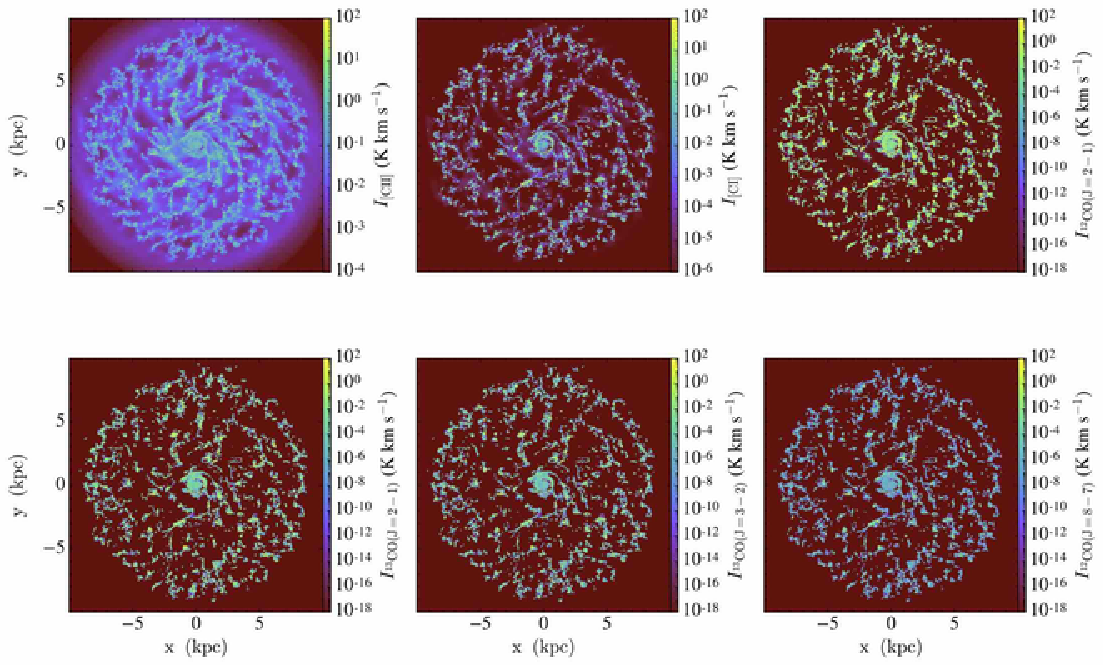}
\caption{
The galactic disk for Run VI (radially variable $G_0$, constant
$\zeta$, 4-pc resolution) at 250 Myr. Images are 20 kpc across. From
\textit{top left} to \textit{bottom right} are shown integrated
intensity maps of: (a) [CII] 158~$\mathrm{\mu m}$; (b) CI
609~$\mathrm{\mu m}$; (c) $^{12}$CO($J$=2--1); (d)
$^{13}$CO($J$=2--1); (e) $^{13}$CO($J$=3--2); (f) $^{12}$CO($J$=8--7).
}
\label{fig:best1}
\end{figure*}

In Figure \ref{fig:best1} we show several diagnostic tracers of the
global disk of Run VI at 250~Myr: [CII] 158~$\mathrm{\mu m}$, CI
609~$\mathrm{\mu m}$, $^{12}$CO($J$ = 2--1), $^{13}$CO($J$ = 2--1),
$^{13}$CO($J$ = 3--2) and $^{12}$CO($J$ = 8--7). Note that the adopted
abundance ratio of $^{13}$C to $^{12}$C is 1/60. In constructing these
maps we have simply summed the local emissivities and not accounted
for optical depth. However, such optical depth effects are partially
accounted for in the local 1D PDR models. This will mostly affect the
$^{12}$CO(2-1) results, which should be regarded as illustrative and
not meant for quantitative assessment of line intensities. However,
the other CO tracers are expected to be closer to being optically thin
and thus more accurately represented by these maps.

As expected, the [CII] 158~$\mathrm{\mu m}$ emission is much more
diffuse than the CO emission, i.e., tracing not only the cooler,
denser structures, but also warmer and lower density regions ($n_\HH
\sim$10 -- 100 ${\rm cm^{-3}}$) that are near the transition from the
atomic to molecular phase.  We see that CO emission lines are emitted
from much more discrete, localized clouds.
$^{12}$CO($J$ = 2--1), $^{13}$CO($J$ = 2--1) and $^{13}$CO($J$ = 3--2)
intensity maps show similar structures. On the scales shown in
Figure~\ref{fig:best1}, $^{12}$CO($J$ = 8--7) also shows similar
morphologies, but with lower overall intensities.

\begin{figure}[htb!]
\centering
\includegraphics[width = 0.5 \textwidth]{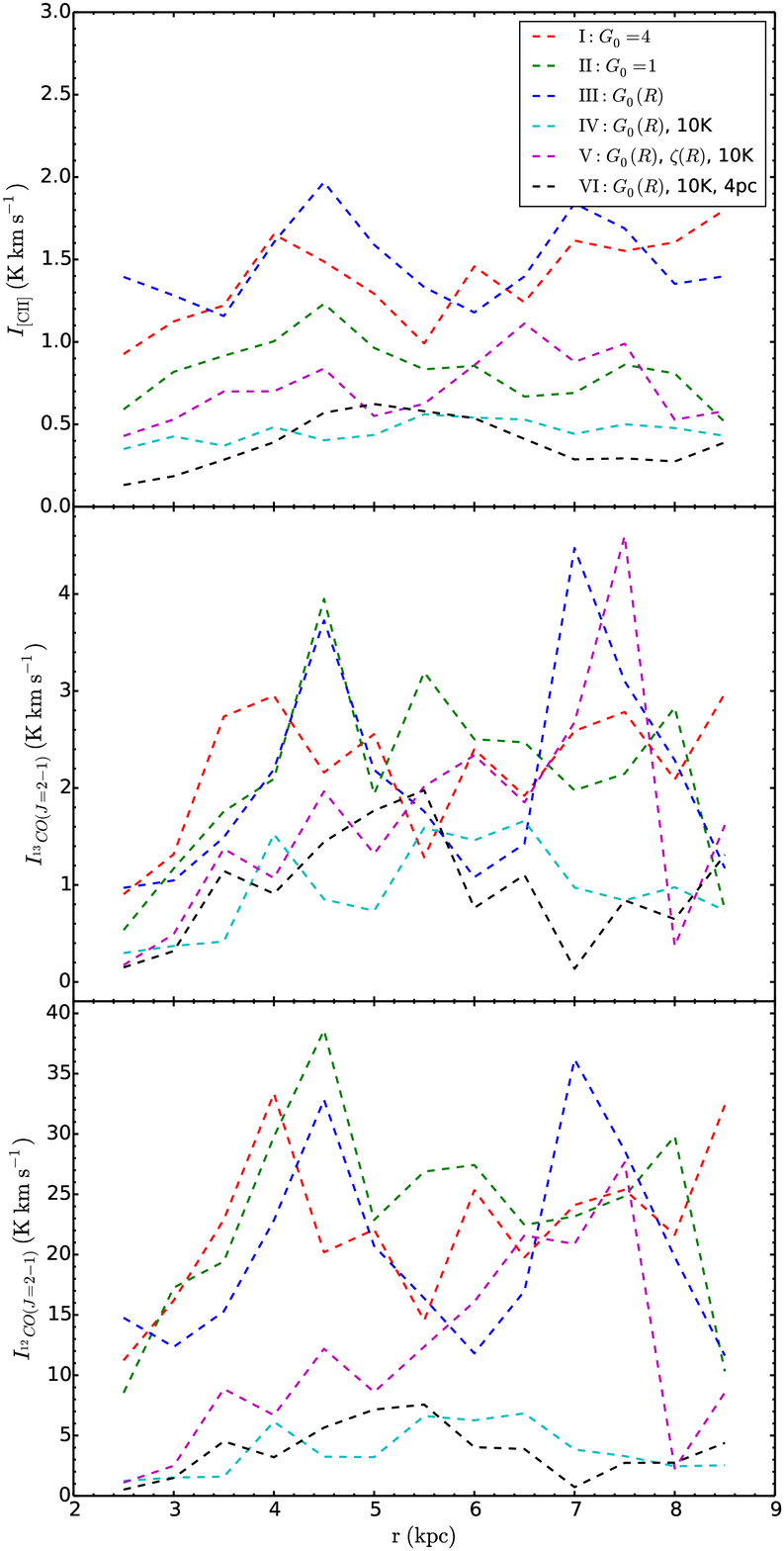}
\caption{
Area-weighted integrated [CII] 158~$\mathrm{\mu m}$ (\textit{top
  panel}), $^{13}$CO($J$ = 2--1) (\textit{middle panel}) and
$^{12}$CO($J$ = 2--1) (\textit{bottom panel}) emission line
intensities along the line of sight as a function of galactocentric
radii for face-on views of the disks of Run I (red), II (green), III
(blue), IV (cyan), V (magenta), VI (black) (all results for
$t=250\:$Myr).}
\label{fig:ciir}
\end{figure}

Figure \ref{fig:ciir} shows the area-weighted integrated [CII] (top
panel), $^{13}$CO($J$ = 2--1) (middle panel) and $^{12}$CO($J$ = 2--1)
(bottom panel) line intensities along the line of sight as functions
of galactocentric radii for face-on views of the disks at
250~Myr. Comparing runs with the same temperature floor we can see
that, overall, a stronger FUV radiation field leads to higher [CII]
intensities, but has less influence on CO lines because it attenuates
significantly through the dense regions where CO lines are
strong. 


\subsection{Example Kiloparsec Patches}

\begin{figure*}[htb!]
\centering
\includegraphics[width = 1.0\textwidth]{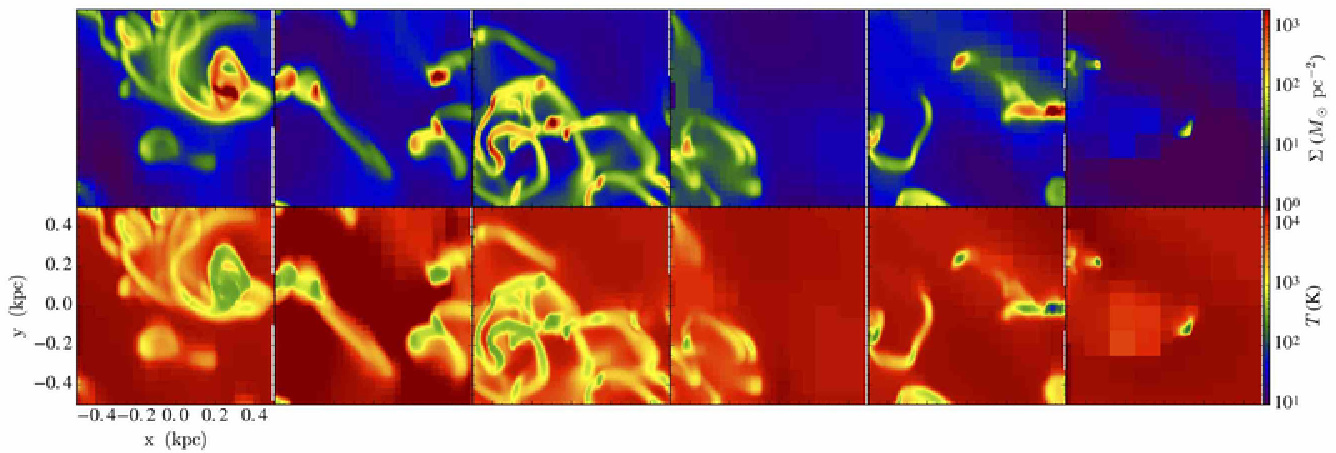}
\caption{
{\it (a) Top row:} The mass surface density structures of 1-kpc-wide
patches of the disks (averaged vertically over a 2~kpc vertical extent
centered on disk midplane) centered at $(x,y)=(4.25,0)$~kpc at
250~Myr, extracted from Runs I to VI (left to right). This part of the
disk is in a quasi-steady state at this time with respect to the
distribution of ISM phases for all these simulation runs, but large
variations still occur on $\sim$kpc scales. {\it (b) Bottom row:} As
(a), but now showing mass-weighted temperature.}
\label{fig:patch1}
\end{figure*}

To examine the structure of the ISM on scales $\lesssim 1\:$kpc, we
extract example 1~kpc disk patches from the simulations at
galactocentric radius of 4.25~kpc and at 250~Myr and visualize their
mass surface density and temperature structures in Figure
\ref{fig:patch1}. These are the locations and times extracted from the
TT09 simulation and used as initial conditions for higher resolution
simulations by \citet{VL13,VL15}, \citet{Butler15} and \citet{Butler17}.

The patches show large variations from run to run, which is caused by
the chaotic nature of small scale ISM, including GMC,
evolution. General features include the presence of elongated,
filamentary structures, some of which may be related to tidal tails
around denser clouds. In the patches where there are high numbers of
dense clouds, there is clear evidence for interactions.

\begin{figure}[htb!]
\centering
\includegraphics[width = 0.5\textwidth]{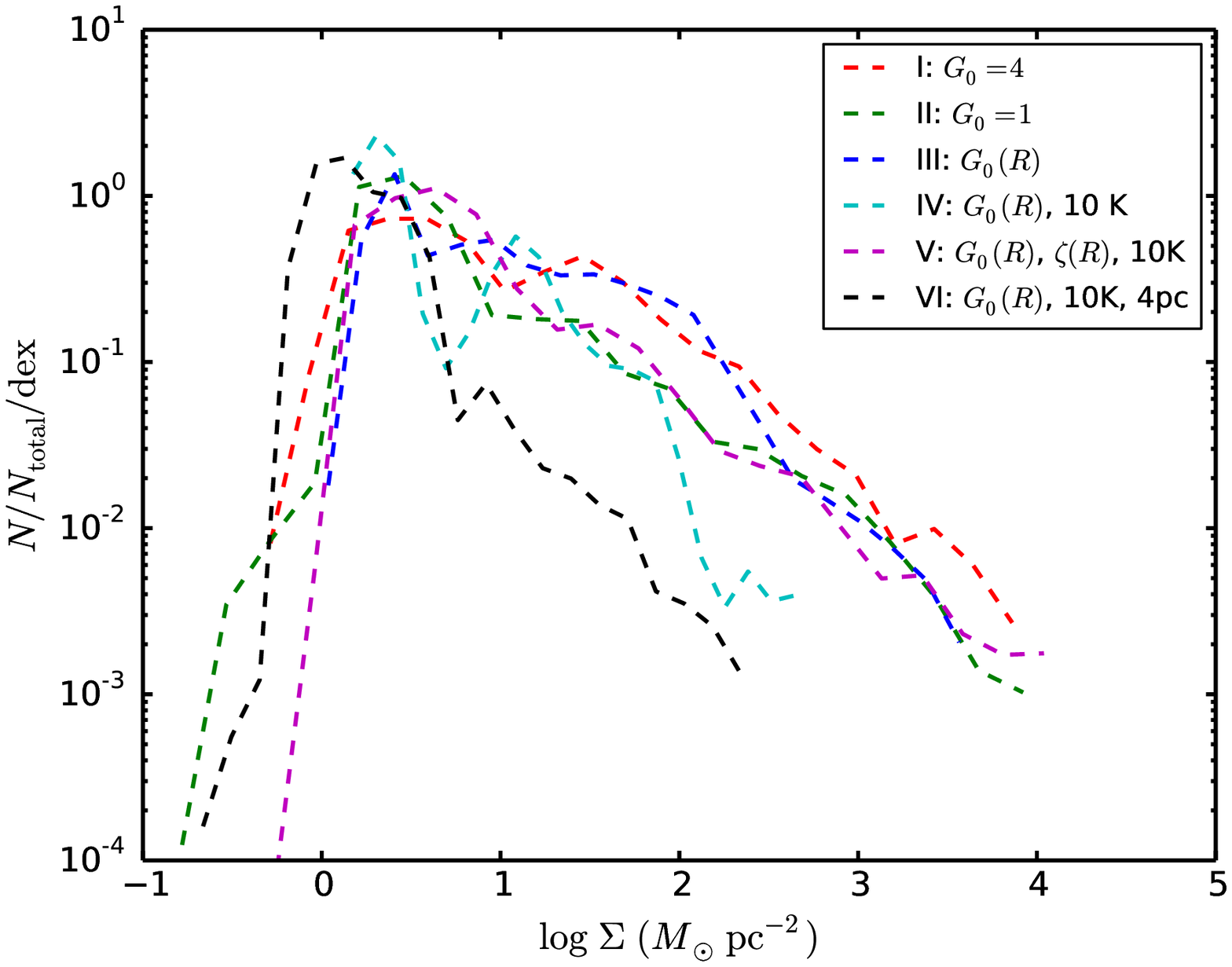}
\caption{
Probability distribution function (PDF) of gas mass surface density
evaluated in the 1~kpc patch regions centered at
$(x,y)=(4.25,0)$~pc at 250~Myr for Runs I (red), II (green), III
(blue), III (blue), IV (cyan), V (magenta) and VI (black).}
\label{fig:patchsigma}
\end{figure}

The probability distribution functions (PDFs) of mass surface density
(calculated with a fixed linear resolution, 8~pc for Runs I to V and 4pc
for Run VI) are shown in Figure~\ref{fig:patchsigma}. Again, these
show a variety of shapes, but generally peak at few $M_\odot\:{\rm
  pc}^{-2}$, with high-end power law tails. A great deal of variation
is due simply to stochastic sampling within a galaxy: for example, the
region extracted from Run VI only has a couple of small GMCs and so
has the smallest areal fraction of material above, e.g.,
$10\:M_\odot\:{\rm pc}^{-2}$.

\begin{figure*}[htb!]
\centering
\vspace{-0.3in}
\includegraphics[width = 0.7\textwidth]{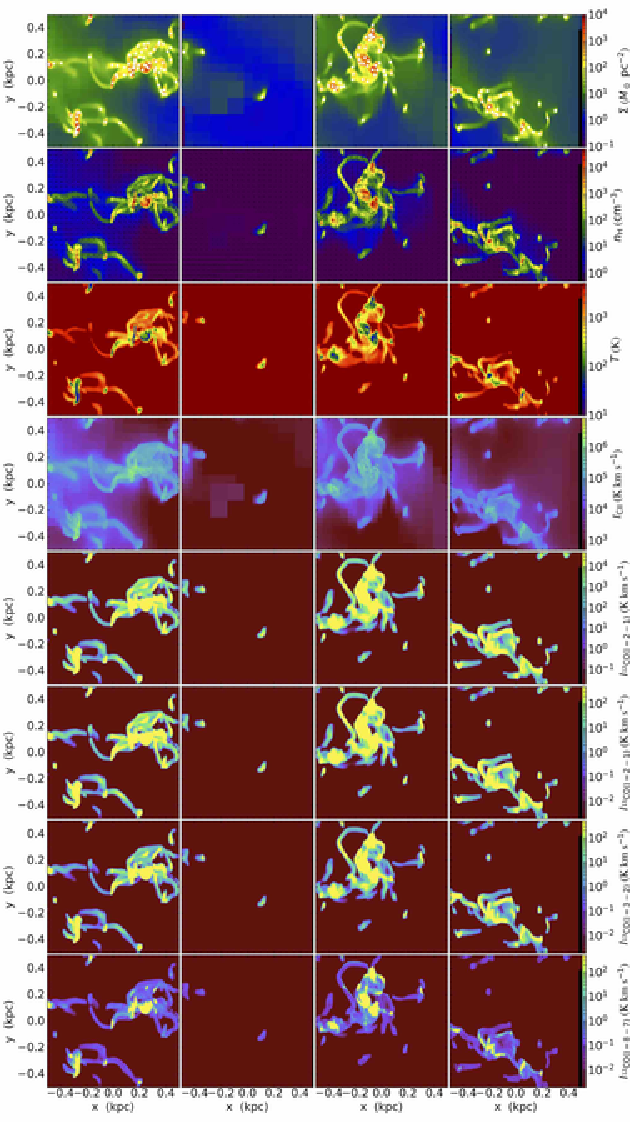}
\vspace{-0.2in}
\caption{
Examples of 1~kpc sized patches (integrated over 2 kpc in the
$z$-direction), extracted from the main disk of Run VI at 250 Myr,
centering at $r = 3$, 4.25, 5.25, and 7~kpc (columns from left to
right) in the mid plane. Rows from top to bottom show the total gas
mass surface density ($\Sigma$), mass-weighted number density of H
nuclei ($n_{\rm H}$), mass-weighted temperature ($T$), and integrated
[CII] 158~$\mathrm{\mu m}$, $^{12}$CO($J$=2--1), $^{13}$CO($J$=2--1),
$^{13}$CO($J$=3--2) and $^{12}$CO($J$=8--7) line intensities. The
center of mass of identified clouds are marked as white points in the
top row, which also shows the local velocity field, i.e., after
subtracting the mean circular orbital velocity at each location.}
\label{fig:best2}
\end{figure*}

We also display a series of kiloparsec patches from Run VI, centering
at $r = 3$, 4.25, 5.25 and 7 kpc (Figure~\ref{fig:best2}).  These
regions contain a great variety of structures, as seen in their mass
surface densities, mass-weighted number densities and temperatures,
and integrated intensities of emission lines [CII] 158~$\mathrm{\mu
  m}$, $^{12}$CO($J$=2--1), $^{13}$CO($J$=2--1), $^{13}$CO($J$=3--2)
and $^{12}$CO($J$=8--7) along the line of sight ($z$-direction). Here
in the $\Sigma$ map we also show the center of mass locations of
identified ``GMCs.'' As noted earlier, [CII] 158~$\mathrm{\mu m}$ line
emission, tracing warmer and lower density ($n_\HH \sim$10--100${\rm
  cm^{-3}}$) gas, is much more diffuse and extended than that of CO
lines.

In principle, the statistics of line and dust continuum emission from
ensembles of such kpc patches, or whole galaxies, may be compared to
observations of nearby galaxies, e.g., with {\it ALMA} or {\it
  SOFIA}. Such data could first be used to constrain a global model
for a galactic disk (e.g., rotation curve, total gas $\Sigma(r)$
profile), then allowing a tailored simulation of a particular
system. Then the detailed statistics of the distributions of
emissivities could be used to test the validity of the simulations,
e.g., comparing the effects of including $B$-fields and/or methods of
implementing feedback. Such comparisons are are a longer term goal of
this project.


\section{GMC Physical Properties}
\label{sec:gmc}

\begin{figure*}[htb!]
\centering
\includegraphics[width = 1.0 \textwidth]{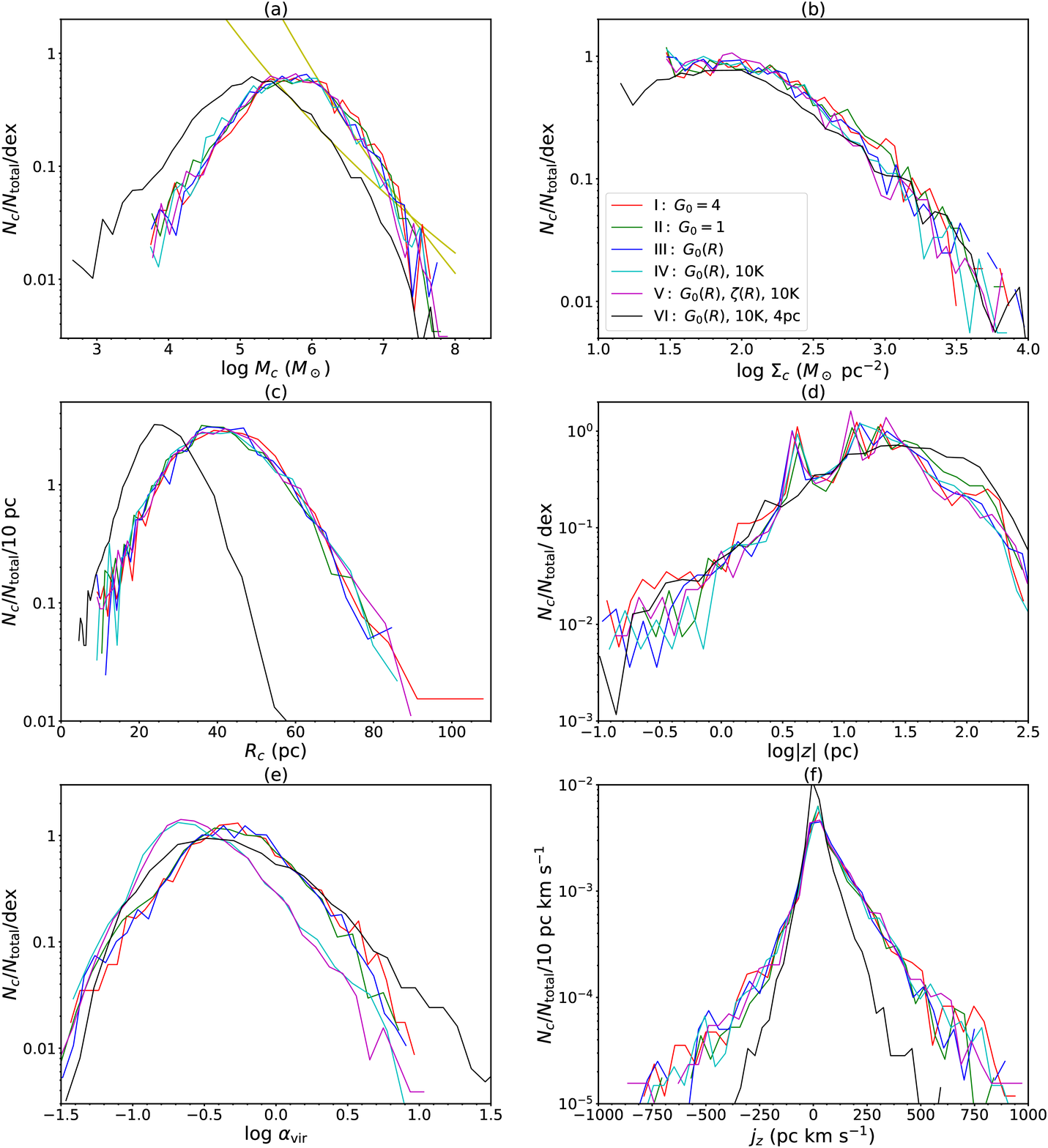}
\caption{
Probability distribution functions (PDFs) of GMC properties for Runs I
(red), II (green), III (blue), IV (cyan), V (magenta) and VI (black)
at 300 Myr (see legend). The panels show: (a) mass; (b) mass surface
density; (c) average radius; (d) center-of-mass vertical position; (e)
virial parameter; (f) specific angular momenta. The straight yellow
lines in (a) show power law fits to the high-end GMC mass spectra
$\diff N_c/\diff \: {\rm log}\: M_c \propto M_c^{-\alpha}$ with
$\alpha=1.43\pm0.08$ and $0.93\pm0.03$ for Run IV and VI,
respectively, at 300 Myr (see text).
}
\label{fig:cc}
\end{figure*}

\begin{figure*}[htb!]
\centering
\includegraphics[width = 1.0 \textwidth]{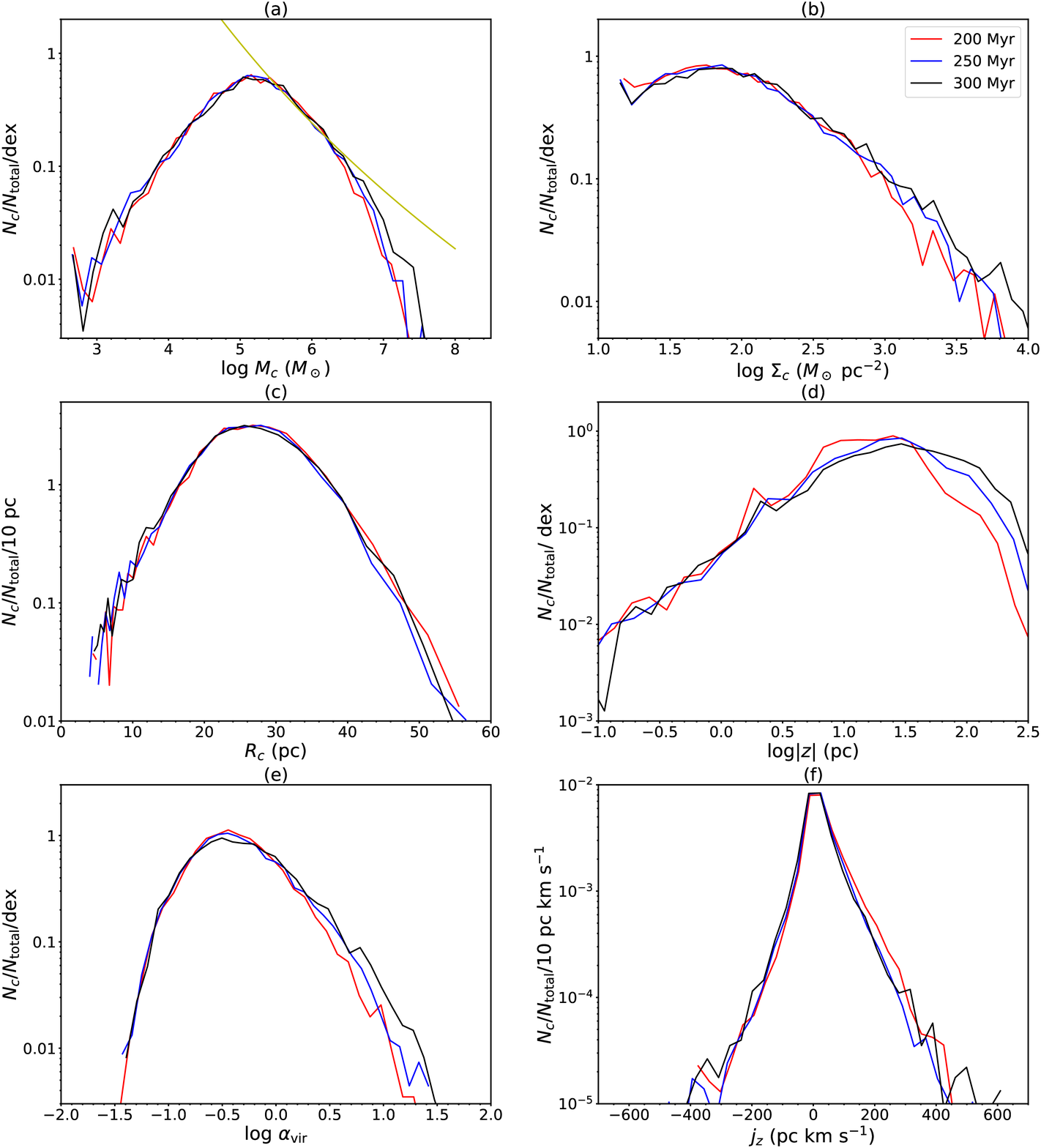}
\caption{
As Figure~\ref{fig:cc}, but now showing results for Run VI at 200, 250
and 300 Myr. The panels show: (a) mass; (b) mass surface density; (c)
average radius; (d) center-of-mass vertical position; (e) virial
parameter; (f) specific angular momenta. The straight yellow line in
(a) show a power law fit to the high-end GMC mass spectrum $\diff
N_c/\diff\: {\rm log}\: M_c \propto M_c^{-\alpha}$ with $\alpha=0.93$ at
300 Myr.}
\label{fig:run6}
\end{figure*}

In this section we show the physical properties of the identified
``GMCs,'' which, recall, are connected, locally-peaked structures
identified by having $n_{\rm H}\geq100\:{\rm cm}^{-3}$ (see
\S\ref{sec:method}). Our analysis methods follow those of TT09. We
focus on GMC properties during the period from 200 to 300 Myrs when
the disks are in quasi-statistical equilibrium in terms of their
global ISM properties, as shown in the previous
section. Figure~\ref{fig:cc} compares the probability distribution
functions (PDFs) of several GMC properties for all six runs at
300~Myr. Figure~\ref{fig:run6} presents the time evolution of these
PDFs from 200 to 300 Myr for Run VI.

GMC mass ($M_c$) PDFs are shown in Figure~\ref{fig:cc}(a). For Runs I
to V (i.e., with 8~pc resolution), these PDFs show a broad peak near
$5\times10^5\:M_\odot$, with clouds seen with masses from
$\sim10^4\:M_\odot$ to $\sim5\times10^7\:M_\odot$. We see that these
GMC mass functions are fairly insensitive to the modeled PDR and CR
physics.
However, the linear resolution of the simulation has a significant
effect: i.e., Run VI has a GMC mass function that peaks at lower
masses. We fit power laws of the form $\diff N_c/\diff \:{\rm log}\:
M_c \propto M_c^{-\alpha}$ to the high-end range ($>10^6\:M_\odot$ for
Runs I to V; $>10^{5.5}\:M_\odot$ for Run VI), with errors set from
Poisson counting statistics in each mass interval. Runs I to V have
values of $\alpha\simeq1.4$, while Run VI has $\alpha\simeq
0.93\pm0.03$. Thus we see that the resolution of the simulation has a
significant effect on the mass function of the GMCs, as defined with
our selection criteria of \S\ref{sec:method}. Note that our selection
of GMCs in Run VI uses the same deblending distance parameter of 32~pc
as the other runs, so the difference in the mass function arises
because of Run VI's ability to better resolve fragmentation of the
structures.

Figure~\ref{fig:run6}(a) shows that the GMC mass function remains very
constant from 200 to 300~Myr for Run VI, with the most noticeable
evolution being a small increase in the number of the most massive
clouds by the time of 300~Myr. As discussed earlier, detailed results
of GMC mass functions are likely influenced by the lack of localized
star formation and feedback in these simulations, since diffuse FUV
feedback on its own is ineffective at destroying GMCs. However, since
the mass fraction of gas in GMCs is quite stable (\S\ref{sec:ism}), we
hypothsize that in these simulations it is the process of GMC
collisions that act to regulate the GMC population, i.e., keeping
$\sim1/3$ of the gas in the diffuse intercloud medium.



We can compare the simulated GMC mass distributions to those observed
in the Milky Way and M33, which can also be fitted with a power law at
their high-mass end. In the Milky Way, $\alpha$ is estimated to be
$\sim 0.6-0.8$ \citep{WM97}, based on analysis of $^{12}$CO(1-0) line
survey data. In the inner ($\leq 2 \:$kpc) region of M33, with
$\sim50\:$pc resolution, \citet{Gratier12} measure
$\alpha\simeq0.6\pm0.2$ (assuming a one to one correspondence of
CO(2-1) line luminosity with GMC mass). At galactocentric distances
$>2\:$kpc, they measure $\alpha\simeq1.3\pm0.2$. While the values of
$\alpha$ seen in Run VI (i.e., $\alpha\simeq 0.93\pm0.03$) are similar
to the above observed values, we caution that the numerical results
are not converged (i.e., with the 4~pc resolution of Run VI, we see a
much shallower high-end power law slope). At the same time, one must
recognize that observational results will also likely depend on the
linear resolution achieved and the particular method of GMC
identification and mass measurement. Most observational studies of GMC
mass functions use $^{12}$CO line emission to identify the clouds,
followed by empirical relations to convert line luminosities to
mass. Since our PDR modeling does not yet include accurate treatment
of the global radiative transfer effects that are important for the
fluxes of optically thick lines, like $^{12}$CO(2-1), we defer more
detailed comparisons of the GMC mass (and line luminosity) functions
to a future paper.


%


Figure \ref{fig:cc}(b) shows the distributions of mass surface
densities, $\Sigma_c\equiv M_c/(\pi R_c^2)$, of the clouds as viewed
along the $z$-axis, perpendicular to the disk, at 300 Myr. Here $R_c$
is the average projected radius of the GMCs in the disk plane,
discussed below. The distributions of mass surface densities are very similar among the
six runs, with a broad peak near $\sim60\:M_\odot\:{\rm pc}^{-2}$ (the
peak becomes most apparent in Run VI), which falls off with a power
law tail at the high $\Sigma$ end. The minimum value of $\Sigma_c$ can
be understood as simply being related to that of a GMC which is only
resolved with a single layer of grid cells in the vertical direction:
i.e., $\simeq 28\:M_\odot\:{\rm pc}^{-2}$ for an 8~pc sided cell
filled with gas at $n_{\rm H}=100\:{\rm cm}^{-3}$ and half this value
for the 4~pc resolution Run VI. On the otherhand, a few, extreme
``GMCs'' have $\Sigma\sim 10^4\:M_\odot\:{\rm pc}^{-2}$. Figure
\ref{fig:run6}(b) shows that the PDFs of $\Sigma$ for Run VI remain
quite constant from 200 to 300~Myr. There is slight growth in the
high-$\Sigma$ tail of the distribution, but this involves only a
relatively small number of clouds.


The mass surface density of the peak of the PDF of the Run VI GMCs is
similar to the mean value of $\sim 200\:M_\odot\:\mathrm{pc}^{-2}$
derived by \citet{S87} from a $^{12}$CO(1-0) survey of the Galaxy.
\citet{Heyer09} have derived smaller values $\sim 100 M_\odot$
$\mathrm{pc}^{-2}$ based on better sampled $^{13}$CO surveys (see
discussion of Tan, Shaske \& Van Loo 2013). Considering the difference
in GMC identifications between observations based solely on molecular
line (CO) emission, and our simulated clouds, $\Sigma$ values of our
simulated GMCs are broadly consistent with observations. Nevertheless,
we again expect that additional physics of $B$-fields, star formation
and localized feedback will all act to reduce $\Sigma$ of the GMC
population. 

Figure \ref{fig:cc}(c) shows the PDFs of average GMC radius,
$R_c\equiv (A_c/\pi)^{1/2}$, where $A_c$ is the projected area viewed
along the $z$ axis. Given the results for the PDFs of $M_c$ and
$\Sigma_c$, the similarity of the PDFs among Runs I to V is not
surprising. Their PDFs peak at $R_c\simeq40\:$pc, then extending up to
$\sim100\:$pc. For Run VI, the peak of the PDF is near $25\:$pc and
the largest GMCs have radii of $\sim50\:$pc. Figure \ref{fig:run6}(c)
shows the constancy of the $R_c$ distribution from 200 to 300~Myr for
Run VI. These results show that the sizes of identified ``GMCs''
depend on the resolution of the simulation and are thus not
numerically converged. Still, the sizes of the clouds in Run VI are
comparable to those observed for GMCs in the Galaxy
(e.g. \cite{MD17}).




Figure \ref{fig:cc}(d) shows the distribution of the magnitudes of the
vertical distances of GMC center of masses from the disk
midplane. Most clouds are located within a narrow range of $\sim30$~pc
from the midplane. The vertical half-scale-height of the population of
GMCs is $|z|_{1/2}=26\:$pc, i.e., the median value of $|z|$. This
half-scale-height is similar to the typical radii of the GMCs,
demonstrating that the GMCs interact as part of a quasi-2D
distribution, which is an assumption of the GMC-GMC collision models
of \citet{G91} and \citet{T00}, discussed below. A small
number of GMCs are at relatively large vertical displacements from the
midplane. Figure~\ref{fig:run6}(d) shows the numbers of these high
$|z|$ GMCs increases moderately from 200 to 300 Myr in Run VI. All
vertical displacements of the GMCs in these simulations are driven by
gravitational forces acting between the self-gravitating
clouds. Generally the largest velocities that will be achieved in such
interactions will be approximately equal to the escape speed from the
clouds. As shown in Fig.~\ref{fig:Q_r}(b), the gas velocity dispersion
in the disk plane can reach 30~km/s at late times in the inner region
of the galaxy. A larger range of $|z|$ at later times is consistent
with the growth of some GMCs to larger masses and higher densities, as
discussed above.

Figure \ref{fig:cc}(e) shows the distribution of GMC virial parameters,
defined as \citep{BertoldiM92}
\begin{equation}
\alpha_{\rm vir} \equiv \frac{5\sigma_c^2 R_c}{G M_c},
\end{equation}
where $\sigma_c$ is the mass-averaged 1D velocity dispersion of the
cloud, i.e., $\sigma_c \equiv (c_s^2+\sigma^2_{\mathrm{nt},c})^{1/2}$
and $\sigma^2_{\mathrm{nt},c}$ is the 1D velocity dispersion about the
center-of-mass velocity of the cloud. A value of $\alpha_{\rm vir}=1$
implies a spherical, uniform cloud with negligible surface pressure
and magnetic fields is virialized, while $\alpha_{\rm vir}<2$
indicates this type of cloud is gravitationally bound. The PDFs of
$\alpha_{\rm vir}$ are relatively broad, but tending to peak at values
$<1$. The effect of the 300~K temperature floor is clearly seen in
Runs I, II and III, since it sets an effective minimum velocity
dispersion of about 1.6~km/s. Runs IV and V have PDFs of $\alpha_{\rm
  vir}$ that peak at significantly lower values. Interestingly, the
distribution of $\alpha_{\rm vir}$ rises again in the GMCs in the
higher resolution simulation Run VI. Figure~\ref{fig:run6} shows this
distribution remains nearly constant from 200 to 300~Myr. At 300~Myr,
its median value is $\alpha_{\rm vir}\simeq0.45$. Most (78\%) of the
GMCs are gravitationally bound, consistent with the observational
results of \citet{RD10}, as analyzed by \citet{Tan13}. GMCs that are unbound are likely due to the dynamical
effects of strong gravitational interactions, including recent
collisions, between GMCs.


Figure \ref{fig:cc}(f) shows the distribution of GMC specific angular
momenta about a rotation axis through the center of mass perpendicular
to the disk plane, i.e., $j_z$. Positive values of $j_z$ indicate
prograde rotation, i.e., in the same sense as the orbital motion of
the galaxy, while negative values indicate retrograde rotation. The
distributions are similar among Runs I to V, showing a modest excess
of prograde over retrograde GMCs. The fraction of retrograde GMCs is
$f_{\rm retro}\simeq0.28$ for these GMC populations. The distrbution
in Run VI is narrower, as expected since GMCs are smaller, and its
value of $f_{\rm retro}\simeq0.36$ is slightly greater than the
results of the other runs. Figure \ref{fig:run6}(f) shows that the Run
VI PDF of $j_z$ does not change significantly during the period from
200 to 300~Myr. As discussed by TT09 (see also \cite{Dobbs08,Dobbs11}),
$f_{\rm retro}$ may be a useful diagnostic of the role of GMC-GMC
collisions in a galaxy population. Large values of $f_{\rm retro}$ may
indicate that GMCs are collisionally evolved.
Our results are consistent with observations showing that retrograde
clouds are nearly as common as prograde ones (\cite{Blitz93};
\cite{Phillips99}; \cite{Rosolowsky03}; \cite{Imara11a};
\cite{Imara11b}).

\section{GMC Collisions}\label{S:collisions}

Shear-driven GMC-GMC collisions have been proposed as a mechanism to
trigger star cluster formation, which could then explain the SFRs of
disk galaxies and circumnuclear starbursts (\cite{T00}; see also TT09;
\cite{Tan10,Tan13,Suwannajak14}). This model assumes
the galactic disk has a Toomre parameter $Q\sim1$ and a significant
fraction, $f_{\rm GMC}\sim0.5$, of its gas mass in self-gravitating
clouds, i.e., ``GMCs,'' which occupy an essentially 2D distribution in
the disk plane (i.e., the scale height of the GMCs is similar to the
size of the clouds). These conditions were met by the disk simulated
by TT09 and are also achieved by the new disk simulations presented in
this paper. Under such conditions, gravitational interactions between
GMCs lead to an increase in the velocity dispersion of the clouds,
which is eventually balanced by damping by GMC collisions \citep{G91}. These ``peculiar motions'' of GMCs with respect to their
local circular velocity cause GMCs to interact with other clouds that
are on orbits with slightly different mean galactocentric radii. For a
population of equal mass GMCs, typical collisions occur between clouds
that have initial mean orbital separations, i.e., collision impact
parameters, of $b\sim1.6r_t$ with a range of $\Delta b \sim r_t$,
where $r_t=(1-\beta)^{-2/3}(2M_c/M_{\rm gal})^{1/3}r$ is the tidal
radius of the clouds, with $\beta \equiv d\: {\rm ln}\: v_{\rm circ}/ d\: {\rm ln}\:
r$, and $M_{\rm gal}$ is the total galactic mass interior to
galactocentric radius $r$.

Under these conditions, the average rate of GMC-GMC collisions is
estimated to be (\cite{T00}; see also \cite{Tan13}):
\begin{eqnarray}
t_{\rm coll} & \simeq & \frac{1}{2} \frac{\lambda_{\rm mfp}}{v_s(b=1.6r_t)}\\
 & \simeq & \frac{1}{2 b^\prime(\Omega - \diff v_{\rm circ}/\diff r){\cal N}_A r_t^2 f_G},\label{eq:tcoll} 
\label{eq:tcoll}
\end{eqnarray}
where $\lambda_{\rm mfp}$ is the mean free path between collisions,
$v_s$ is the shear velocity in the disk for orbits separated by a
given initial impact parameter, i.e.,
\begin{equation}
v_s = b (\Omega - \diff v_{\rm circ} / \diff r),\label{eq:vs}
\end{equation}
$b^\prime \equiv b/r_t\simeq 1.6$, $\Omega=v_{\rm circ}/r = 2\pi/t_{\rm orb}$,
${\cal N}_A$ is the number of GMCs per unit area, and $f_G\sim 0.5$ is
the fraction of strong gravitational encounters that lead to
collision. This fraction depends on the physical size of the GMCs and
is somewhat uncertain. For ${\cal N}_A\simeq f_{\rm GMC}
\Sigma_g/\bar{M}_c$ and $\sigma_g\simeq
(G\bar{M}_c\kappa)^{1/3}(1.0-1.7\beta)$ \citep{G91}, we have
\begin{eqnarray}
t_{\rm coll} & \simeq & \frac{0.15 Q t_{\rm orb}}{b^\prime f_{\rm
  GMC} f_G (1-0.7\beta)}\\
 & \rightarrow & 0.19 t_{\rm orb}/f_{\rm GMC},
\end{eqnarray}
where the last evaluation is for constant $v_{\rm circ}$, i.e. $\beta=0$, and
for $f_G=0.5$. If most star formation is initiated by GMC collisions,
then this result predicts $\Sigma_{\rm SFR}\propto \Sigma_g \Omega
(1-0.7\beta)$, i.e., a modification of the ``dynamical''
Kennicutt-Schmidt relation observed by \citet{K98}, and further
tested in resolved nearby galaxies by \citet{Leroy08}, \citet{Tan10}
and \citet{Suwannajak14}.

By tracking GMCs in global galactic disk simulations, TT09 measured
$t_{\rm coll}/t_{\rm orb}$ as a function of $r$, finding typical
values of $\simeq 0.2$, with a tendency for this ratio to decrease
gradually with increasing $r$. GMCs with $M_c>10^6\:M_\odot$ had
$t_{\rm coll}/t_{\rm orb}\sim0.1$; note $t_{\rm coll}$ is the average
time between collisions of these massive GMCs with all other
GMCs. Note also that in the simulations there are effects of there
being a range of GMC masses, as well as collective interactions, i.e.,
in and around GMC complexes, which are not accounted for in the simple
analytic estimates that are based on idealized binary collision
rates. There can also be systematic variations of ${\cal N}_A$ and
$r_t$ with location in the simulated disk galaxy. To account for
these, \citet{Tan13} tested the prediction of equation
(\ref{eq:tcoll}) directly in annuli in the TT09 disk, finding good
agreement: in particular, the radial gradient in $t_{\rm coll}/t_{\rm
  orb}$ was accounted for by systematic variations in ${\cal N}_A$ and
$r_t$ (e.g., partly because of variations in the average GMC mass
$\bar{M}_c$ with galactocentric radius), and the faster collision rate
of more massive clouds due to their larger tidal radii.

In this section we carry out a similar analysis of the GMC collisions
in our simulated disks, as well as examining basic properties of the
collisions, including mass ratios and relative velocities. We also
examine the injection rate of ``turbulent momentum'' into GMCs that
may help explain the maintenance of their internal turbulence and thus
their overall structural properties.

\subsection{Mass Ratios of Colliding GMCs}

\begin{figure}[htb!]
\centering
\includegraphics[width = 0.5 \textwidth]{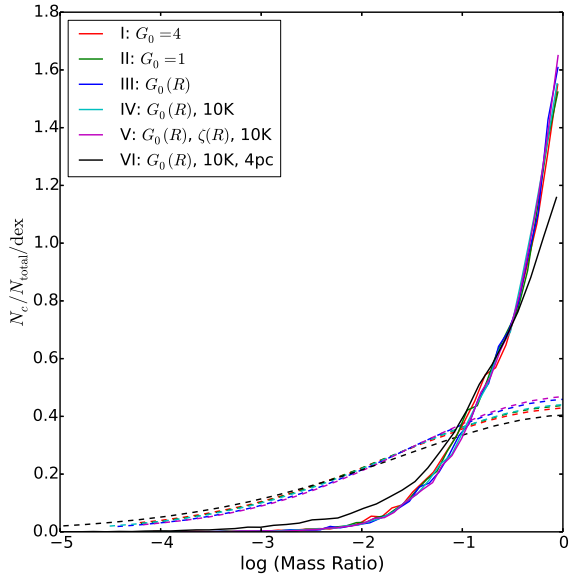}
\caption{
PDFs of mass ratios of colliding GMCs in Runs I to VI from 200 to
300~Myr (solid lines; see legend). The dashed lines with corresponding
colors show the PDFs of mass ratios of randomly selected pairs of
GMCs.}
\label{fig:ccmr}
\end{figure}

We show the PDF of mass ratios ($M_2/M_1$, where $M_2$ is defined as
the smaller mass) of colliding GMCs in Figure~\ref{fig:ccmr} for Runs
I to VI during the simulation time 200--300~Myr. We see that the
typical mass ratio of colliding GMCs peaks close to 1 in all the
simulations, i.e., collisions can be described as ``major mergers.''
This PDF is not a natural result of the cloud mass functions shown in
Figure~\ref{fig:cc}a, since the mass ratio of randomly selected pairs
of GMCs is much broader (Fig.~\ref{fig:ccmr}). Instead, it is caused
by the collision rate to be higher for more massive GMCs, given their
larger tidal radii.

\subsection{GMC Collision Speeds}

\begin{figure}[htb!]
\centering
\includegraphics[width = 0.5 \textwidth]{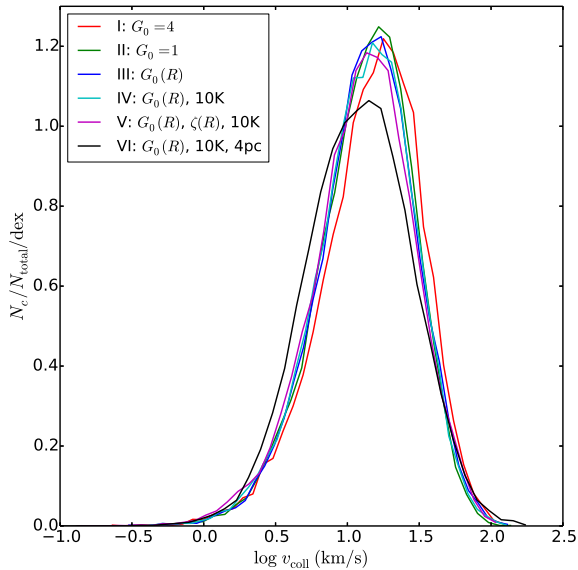}
\caption{
PDFs of relative speeds, $v_{\rm coll}$, of colliding GMCs in Runs I to
VI (see legend) evaluated from 200 to 300 Myr.}
\label{fig:ccrv}
\end{figure}

The PDFs of the relative speeds $v_{\rm coll}$ of colliding GMCs are shown in
Figure~\ref{fig:ccrv}. These distributions peak at $\simeq 17\:{\rm
  km\:s}^{-1}$ for Runs I to V and $\simeq 13\:{\rm km\:s}^{-1}$ for
Run VI. These results are consistent with the collision speeds being
set by the shear velocity from encounters with initial impact
parameters of $b \sim 1$ to 2~$r_t$, i.e., following
equation~(\ref{eq:vs}). The smaller speeds in Run VI would then be
caused by GMCs being of lower mass in this better resolved simulation.
Our results are comparable to recent observations on cloud 
collision (e.g. \citet{Fukui15}).

\begin{figure*}[htb!]
\centering
\includegraphics[width = 1.0 \textwidth]{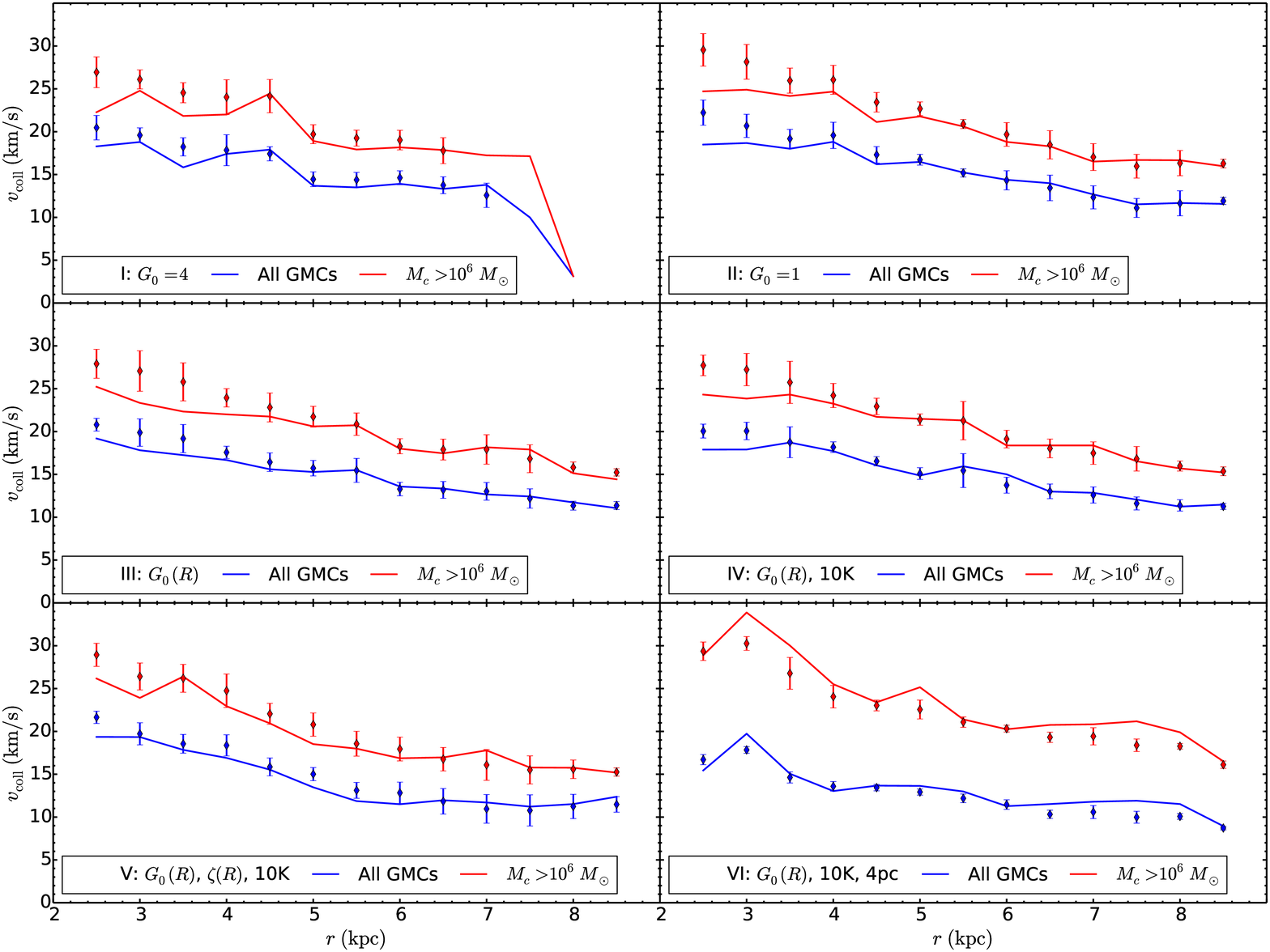}
\caption{
Average relative speeds of colliding GMCs in Runs I to VI from 200 to
300~Myr as a function of galactocentric radius. Blue solid lines show
measured speeds in collisions of all GMCs; red solid lines show the
speeds of collisions where one of the GMCs has a mass
$>10^6\:M_\odot$. The diamonds show analytic estimates of the
collision speeds following equation (\ref{eq:vcoll}), which is based
on the shear velocity at an impact parameter of $b = 1.6 r_t$ plus
$\bar{v}_{\rm ff}/6$ (see text), with the error bar showing $1\sigma$
dispersion in these estimated based on GMC properties from 200 to
300~Myr.
}\label{fig:ccrvr}
\end{figure*}

In Figure \ref{fig:ccrvr} we show the mean relative speed of GMC
collisions as a function of galactocentric radii, averaging in
annuli. The relative speed is seen to be typically a slowly decreasing
function of radius, with values of $\sim10$ to 20$\:{\rm km\:s^{-1}}$
in Run VI when averaging over all GMCs and from $\sim15$ to 30$\:{\rm
  km\:s^{-1}}$ for collisions involving the $>10^6\:M_\odot$
GMCs. These results, along with those of the colliding cloud mass
ratios, can help to guide the initial conditions for numerical
simulations of individual cases of GMC collisions (e.g., \cite{W15,W17a,W17b}).

We compare these speeds to the prediction of $v_{\rm
  coll}=v_s(b=1.6r_t)$, i.e., see equation~(\ref{eq:vs}). This is
generally accurate to within $\sim10\%$, but does systematically
underestimate the collision speed. Gravitational acceleration is
expected to boost the collision speed by an amount of up to of order
the free-fall speed, $v_{\rm ff}$, though geometric effects and gas
pressure gradients will lower the amount of the increase. We find that
a collision speed given by
\begin{equation}
v_{\rm coll} = v_s(b=1.6r_t) + (1/6) \bar{v}_{\rm ff},
\label{eq:vcoll}
\end{equation}
where $\bar{v}_{\rm ff}=(2G\bar{M}_c/\bar{R}_c)^{1/2}$, gives an
accurate description of the mean collision speeds in Runs I to VI and
for collisions between all GMCs and for those involving GMCs with
$>10^6\:M_\odot$ (see Fig.~\ref{fig:ccrvr}).

\subsection{GMC Collision Timescales}

\begin{figure*}[htb!]
\centering
\includegraphics[width= 1.0 \textwidth]{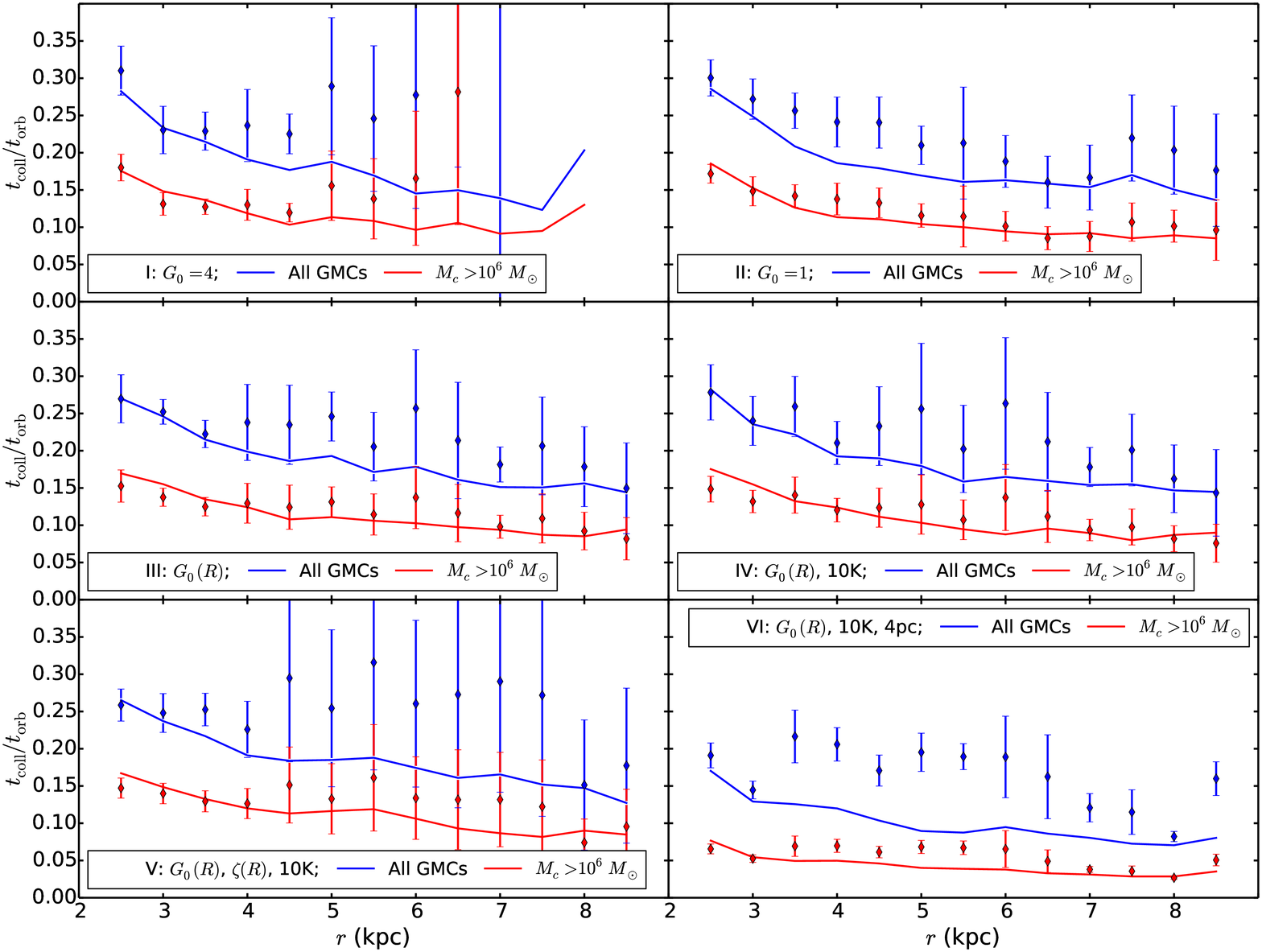}
\caption{
Cloud collision timescales (measured from 200 to 300 Myr) relative to
orbital timescale as a function of galactocentric radius for Runs I to
VI (panels as labelled) averaged for all GMCs (blue solid lines) and
for GMCs with $M_c> 10^6 M_\odot$ (red solid lines).
Diamonds show the theoretical estimates from equation~\ref{eq:tcoll},
averaged for GMC populations from 200 to 300 Myr, with the error bars
showing the $1\sigma$ standard deviation (see text).}
\label{fig:gmccolltime}
\end{figure*}

Figure \ref{fig:gmccolltime} shows the ratio of mean GMC collision
time to local orbital time, $t_\mathrm{coll}/t_{\rm orb}$, measured in
simulation Runs I to VI from 200--300 Myr as a function of
galactocentric radius averaged over all GMCs (blue lines) and for GMCs
with $>10^6\:M_\odot$ (red lines). Note the local orbital time is
given by
\begin{equation}
t_\mathrm{orb} = 123(r/4 \kpc)(v_{\rm circ}/200\mathrm{km\ s^{-1}})^{-1}\ \mathrm{Myr}.
\end{equation}
The average values of $t_\mathrm{coll}/t_{\rm orb}$ are measured by
counting the total number of GMC collisions and the average total
number of GMCs in annuli in the 200 to 300~Myr period, recognizing
that each collision involves two GMCs, and using the value of $t_{\rm
  orb}$ appropriate for the middle of each annulus. We also calculate
the theoretically expected value of $t_\mathrm{coll}/t_{\rm orb}$
based on equation~(\ref{eq:tcoll}) for $t_{\rm coll}$, which assumes
$f_G=0.5$ and uses the number of GMCs per unit area and the mean value
of GMC tidal radii, i.e., depending on $M_c$ and $r$, as inputs. These
estimates are made for the GMC populations measured in the simulation
outputs every 1~Myr from 200 to 300~Myr, with the average results
shown as diamond shaped points for all GMCs (blue) and
$>10^6\:M_\odot$ GMCs (red), with the vertical lines showing the
$1\sigma$ dispersions.

Similar to the results of TT09 and \citet{Tan13}, the mean
collision times averaged for all GMCs are seen to be a small fraction,
$\sim 0.2$, of an orbital time for Runs I to V, with a gradual decline
in this ratio with increasing galactocentric radius. The more massive,
$>10^6\:M_\odot$ GMCs also follow this trend, but have average values
of $t_{\rm coll}/t_{\rm orb}\sim 0.1$. In the higher resolution Run VI
these general results are also seen, but with collision timescales
that are shorter by factors of about $\sim0.5$. The predictions of the
two-body collision rate of GMCs in a shearing disk are quite accurate,
especially for the more massive GMCs. The largest deviation is seen in
Run VI for the ``all GMC'' case, where the observed timescale of
collisions in the simulation can be about a factor of 0.5 shorter than
the prediction. We suspect that this may be due to the GMCs being
organized into larger complexes, where collisions are more frequent,
with such collective effects being less apparent in the lower
resolution runs and when only considering the most massive GMCs.

In summary, at typical locations in the disk of $\sim4\:$kpc, average
times between GMC collisions are only $\sim10$ to 20~Myr, with rates
in approximate agreement with a simple model of 2-body interactions
set by shear in a thin disk \citep{G91,T00}.
These values are relatively short compared to traditional estimates of
cloud collision timescales of hundreds of Myr (e.g. \cite{MO07}).
The difference arises because these traditional estimates consider a
3D (rather than 2D) geometry, invoke collision velocities set by GMC
velocity dispersions (rather than shear velocities at one to two tidal
radii), and have smaller collision cross sections using actual GMC
sizes (rather than GMC tidal radii that results from gravitational
focussing). The general implications of frequent GMC collisions are
discussed, below, starting with their ability to maintain turbulent
motions within GMCs.

\subsection{Maintenance of GMC Turbulence}\label{S:turb}

As reviewed by \citep{MO07}, observed GMC velocity
dispersions are highly supersonic. However, from the results of
numerical experiments, supersonic turbulence is expected to decay in
$\sim$1 dynamical time, $t_{\rm dyn} = R_c/\sigma_c$, where $\sigma_c$
is the 1D internal velocity dispersion \citep{Stone98,ML98} (for $\alpha_{\rm vir}\simeq 1$, $t_{\rm ff} \simeq 0.5
t_{\rm dyn}$). Thus maintenance of turbulence is a constraint on
models of GMC formation and evolution.

\citet{MO07} discussed two conceptual frameworks. First, GMCs are
dynamic, transient and largely unbound, with turbulence driven by
large-scale colliding atomic flows that form the clouds (e.g.,
\cite{Heitsch05,VS11,II12}). These models do not naturally explain why
most GMCs are bound with $\alpha_{\rm vir}\sim 1$. More recent
simulations with moderate strength $B$-fields show that it is
difficult to compress the gas to observed GMC densities (e.g.,
\cite{II08,KB15}). Also, the required fast flows of HI around GMCs
needed to form them quickly in $\sim 1 t_{\rm dyn}$ have not been
observed. For example, \citet{Bihr15} find approximately equal masses
of H$_2$ and HI in the W43 region out to radii of $\sim 140\:$pc. Thus
formation of the GMC in $\sim 1t_{\rm ff}$, i.e., $\sim10\:$Myr, would
require this HI to be converging radially on the W43 GMC location at
$\sim10\:{\rm km\:s}^{-1}$, which is an unlikely initial condition for
the state of the gas before the GMC existed. Similarly,
\citet{Rebolledo17} find that the atomic mass around the Carina Nebula
Gum 31 region (out to $\sim70\:$pc) is only about 1/3 of the total
(although optical depth corrections may provide a modest boost).
In other words, locally the mass fractions of gas that are in these
GMCs are too large compared to the reservoir of HI, since in these
scenarios one does not expect a very high efficiency of conversion of
HI into a GMC. Similarly, it is unclear if such models are globally
consistent with the relatively high mass fractions of gas in GMCs, at
least inside the solar circle (e.g., \cite{Koda16}).

The second framework is that GMCs form by large-scale gravitational
instabilities and are thus gravitationally bound with $\alpha_{\rm
  vir}\sim 1$. Turbulence is maintained by contraction and then,
later, star formation feedback (e.g., \cite{GK11}). While
this paradigm may be relevant in regions of galaxies that are HI
dominated, we argue below that in regions where a large fraction of
gas is already organized into self-gravitating GMCs and their atomic
envelopes, then other processes become more important for controlling
GMC dynamics and evolution.

A third paradigm of GMC evolution mediated by frequent GMC-GMC
collisions has been presented by \citet{T00}.  The implications of
this scenario for the maintenance of turbulence by frequent collisions
has been discussed by \citet{Tan13}. The rate of injection of
``internal turbulent momentum'' per GMC is $\dot{p}_{\rm CC} =
\bar{p}_{\rm CC}/\bar{t}_{\rm coll}$, where $\bar{p}_{\rm CC}$ is the
average internal turbulent momentum injected into a post-collision
GMC. Note $\bar{p}_{\rm CC}$ is treated as a scalar quantity. It is
evaluated by considering that each collision between two GMCs involves
a certain amount of momentum being deposited into the resulting GMC,
i.e., $p_{\rm CC} = M_1 v_1 + M_2 v_2$, where $v_1$ and $v_2$ are the
pre-collision velocities of the GMCs in the center of mass frame of
the final cloud. Thus in the case of equal mass colliding GMCs,
$p_{\rm CC}= 2 M_c v_{\rm coll}/2 = M_c v_{\rm coll}$. Note, that the
pre-collision GMCs are expected to have a high degree of internal
substructure, coupled together with moderately strong $B$-fields, and
so we are assuming the result of the collision is mostly a high degree
of unbalanced shocks that generate internal turbulence in the
post-collision cloud. Thus, in the limit of high efficiency of
conversion of bulk collision momentum into internal turbulent
momentum, we have
\begin{eqnarray}
\dot{p}_{\rm CC} & = & \bar{M}_c \bar{v}_{\rm coll}/ \bar{t}_{\rm coll}\nonumber\\
 & = & \frac{b^\prime G^{1/3}}{2^{2/3}\pi \bar{f}_{\rm coll}}\left(\frac{\bar{M}_c v_{\rm circ}}{r}\right)^{4/3},\label{eq:pdot}
\end{eqnarray}
where in the second line we have assumed the case of a flat rotation
curve and where $f_{\rm coll}\equiv t_{\rm coll}/t_{\rm orb}$.

\begin{figure*}[htb!]
\centering
\includegraphics[width= 1.0 \textwidth]{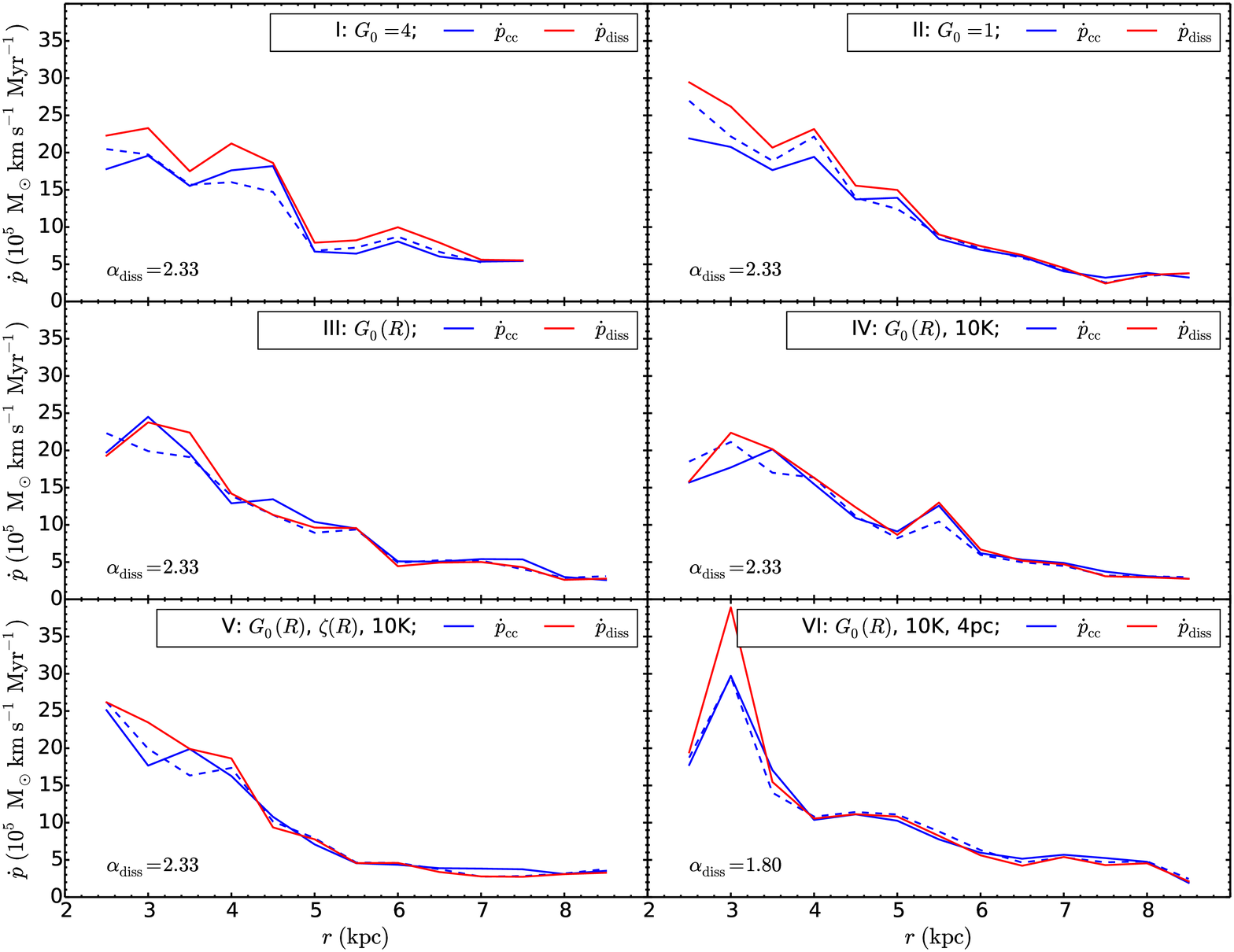}
\caption{
Average rate of turbulent momentum injection per GMC, $\dot{p}_{\rm
  CC}$, as a function of galactocentric radius, measured from 200 to
300 Myr by summing over all recorded GMC collisions, for Runs I to VI
(panels as labelled) (blue solid lines). The prediction from
eq.~(\ref{eq:pdot}) is also indicated (blue dashed line). Also shown in the
panels are estimates of the rate of dissipation of turbulent momentum
per GMC, $\dot{p}_{\rm diss}$, due to internal shocks (red solid
lines). The coefficient for this dissipation rate, $\alpha_{\rm
  diss}$, has been normalized separately for Runs I to V and for Run
VI in order to give an approximate balance $\dot{p}_{\rm diss}\simeq
\dot{p}_{\rm CC}$, with the values as indicated in the panels. The
similarity of the radial profiles indicates that GMC collisions are
driving internal turbulence in the GMCs, balancing the internal decay
rate (see text).}
\label{fig:pdot}
\end{figure*}

In Figure~\ref{fig:pdot} we show the radial profiles of $\dot{p}_{\rm
  CC}$ in the simulations, evaluated from 200 to 300~Myr and averaged
over all GMCs. As expected from our previous results, the momentum
injection rate per GMC is highest in the inner galaxy. It is then seen
to decline by about a factor of $\sim 5$ as one progresses outwards to
$\sim 8\:$kpc.

The rate of dissipation of the total internal turbulent momentum,
$p_c$, of a GMC can be expressed as \citep{Tan13}:
\begin{eqnarray}
\dot{p}_{\rm diss} & \equiv & - p_c / (\alpha_{\rm diss}t_{\rm
  ff}) =  -\sqrt{3}\bar{M}_c\sigma_c/(\alpha_{\rm diss}t_{\rm ff})\nonumber\\
&  = & -2.19
\bar{\alpha}_{\rm vir}^{1/2}G\bar{M}_c\bar{\Sigma}_c/\alpha_{\rm diss},\label{eq:pdiss}
\end{eqnarray}
where $\alpha_{\rm diss}$ is a dimensionless constant. The turbulent
energy dissipation rate has been measured in numerical simulations to
be $\epsilon_{\rm diss} =
(1/2)(\dot{E}/{E})[l_0/(\sqrt{3}\sigma)]\simeq 0.6$ (see review by
\cite{MO07}), so that $\alpha_{\rm diss}\simeq 3.85$ (in a cloud with
$\alpha_{\rm vir}\simeq 1$ and $l_0=2R$). Figure~\ref{fig:pdot} also
shows the radial profiles of $\dot{p}_{\rm diss}$, but where we have
allowed $\alpha_{\rm diss}$ to vary (fixing a single value for Runs I
to V, and another value for the higher resolution Run VI) to obtain a
close match between $\dot{p}_{\rm diss}$ and $\dot{p}_{\rm CC}$. This
requires values of $\alpha_{\rm diss}\simeq 2$, i.e., a shorter
dissipation timescale that is about half the theoretical expectation
of $\simeq 4t_{\rm ff}$, which is not unexpected given the fact the
GMCs in our simulations are only quite poorly resolved. Alternatively,
agreement could be achieved at lower values of $\dot{p}$ by reducing
the efficiency of conversion of the bulk collision momentum to
internal turbulent momentum to about 50\% in equation~(\ref{eq:pdot}).
However, the most important point of Figure~\ref{fig:pdot} is that the
radial profiles of $\dot{p}_{\rm diss}$ and $\dot{p}_{\rm CC}$ are
very similar, suggesting there is a physical link between them.

Finally, we balance the momentum injection rate due to cloud
collisions, given by eq.~(\ref{eq:pdot}), with the dissipation rate
within the clouds, given by eq.~(\ref{eq:pdiss}), to predict typical
GMC equilibrium mass surface densities, $\bar{\Sigma}_{c,{\rm
    eq}}$. 
\begin{eqnarray}
\bar{\Sigma}_{c,{\rm eq}} & = & 102 \bar{\alpha}_{\rm vir}^{-1/2} \frac{\alpha_{\rm diss}}{2} \frac{b^\prime}{1.6}\frac{0.2}{f_{\rm coll}} \left(\frac{\bar{M}_c}{10^6\:M_\odot}\right)^{1/3}\nonumber\\
& \times & \left(\frac{v_{\rm circ}}{200\:{\rm km\:s^{-1}}} \frac{4\:{\rm kpc}}{r}  \right)^{4/3} M_\odot\:{\rm pc^{-2}}.\label{eq:Sigma_eq}
\end{eqnarray}

\begin{figure*}[htb!]
\centering
\includegraphics[width= 1.0 \textwidth]{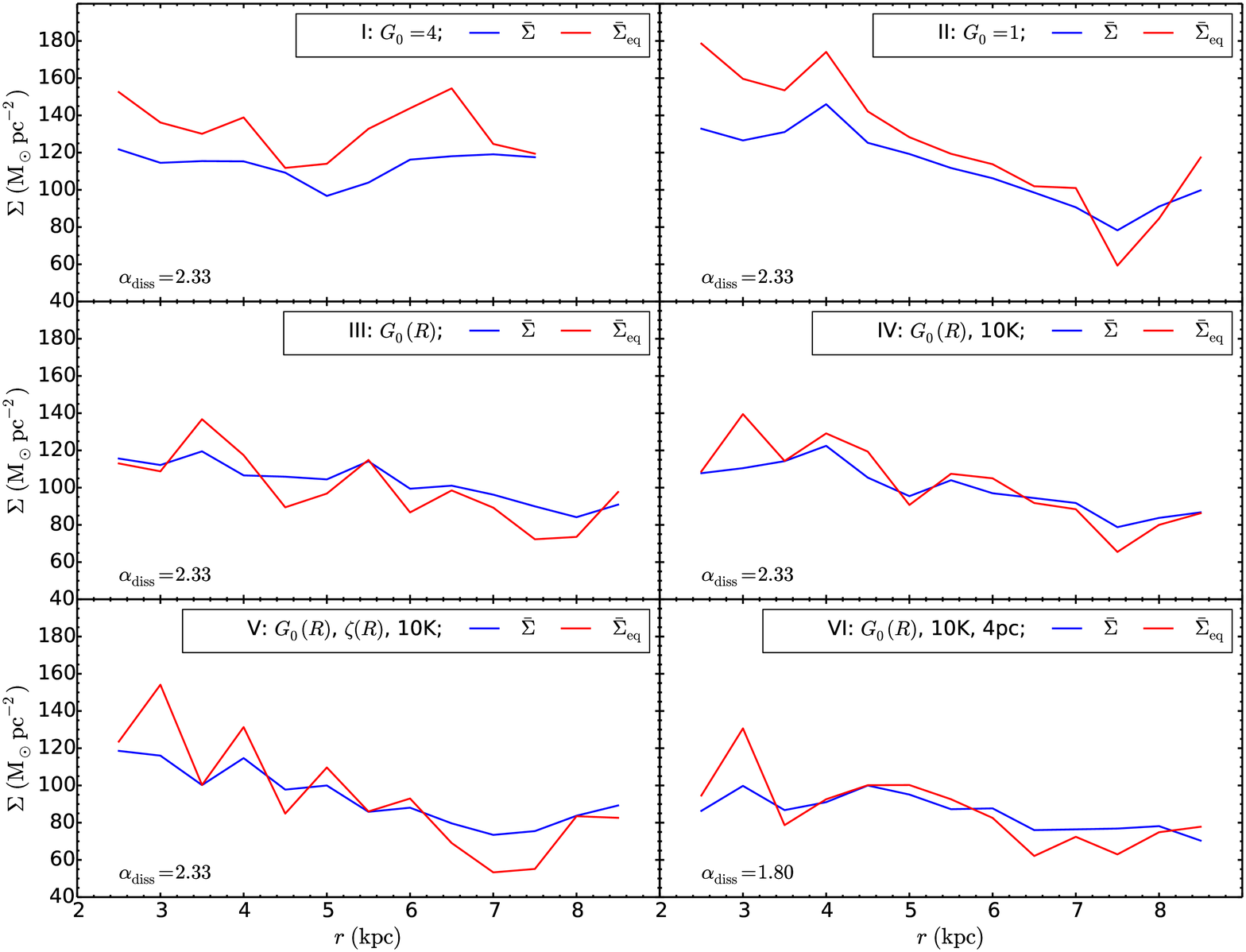}
\caption{
Predicted mean GMC mass surface densities, $\bar{\Sigma}_{\rm eq}$
(red solid lines), measured from 200 to 300 Myr and shown as a
function of galactocentric radius for Runs I to VI (panels as
labelled). This estimate is based on a model of a balance between the
momentum injection rate by GMC-GMC collisions and the internal
dissipation rate resulting from supersonic turbulence, parameterized
by $\alpha_{\rm diss}$ (see text). The actual mean GMC mass surface
densities, $\bar{\Sigma}_c$ (blue solid lines), are also shown. The
similarity of the radial profiles of $\Sigma$ of the GMCs indicates
that their structure can be explained by a simple model in which their
internal turbulence is driven by GMC collisions and decays at a rate
that is inversely proportional to the local free-fall time.
}
\label{fig:Sigma_from_pdot}
\end{figure*}

Figure~\ref{fig:Sigma_from_pdot} shows the predicted radial dependence
of $\bar{\Sigma}_{c,{\rm eq}}$ from eq.~(\ref{eq:Sigma_eq}), which is
calculated by assessing the values of $\bar{M}_c$, $\bar{\alpha}_{\rm
  vir}$ and $\bar{f}_{\rm coll}$ in each annulus. The figure also
compares these estimates with the actual values of the mass surface
densities of the GMCs in the simulations. Since the values of
$\alpha_{\rm diss}$ have been normalized so that $\dot{p}_{\rm
  diss}\simeq\dot{p}_{\rm CC}$, we expect that $\bar{\Sigma}_{c,{\rm
    eq}}$ will on average be approximately equal to $\bar{\Sigma}_c$,
as is seen to be the case. Still, the close correspondence of the
radial profiles of $\bar{\Sigma}_{c,{\rm eq}}$ and $\bar{\Sigma}_c$,
i.e., for a single value of $\alpha_{\rm diss}$, indicates that the
average structural properties of these self-gravitating GMCs are being
set by the injection of turbulent momentum from GMC collisions,
balanced by its rate of decay at a rate inversely proportional to the
local free-fall time.

The equilibrium condition described by eq.~(\ref{eq:Sigma_eq}) is
expected to be prone to instability, since as GMCs become denser and
smaller, they dissipate their turbulent momentum at a faster absolute
rate given their shorter free-fall times. At the same time, being
smaller, they are expected to be less likely to under collisions. In
reality, the processes of $B$-field support, star formation and
feedback are expected to also play a role in regulating the structure
of these densest, collapsing GMCs. In the current simulations, lacking
these processes, we expect that numerical diffusion and viscosity due
to the fast, supersonic motion of the GMCs in their orbits in the
galactic potential leads to some effective support of GMCs that become
too small. In this way, a quasi-equilibrium distribution of GMC
properties can still be established in these simulations (see also
TT09).

\subsection{General Implications of Short GMC Collision Timescales}

Short GMC collision timescales have several important
implications. First, we have already seen in \S\ref{S:turb} that
frequent GMC collisions could be an important driver of turbulence
within the clouds. This mechanism could then regulate the structure of
GMCs, i.e., setting their mass surface densities. The relative
importance of GMC collisions and star formation feedback at driving
internal turbulence needs to be assessed by future studies.

Second, GMC collisions may set the angular momentum distribution of
the clouds, in particular helping to explain the relatively high
fraction of retrograde rotation directions
\citep{Koda06,Imara11a,Imara11b} (see also Fig.~\ref{fig:cc}f and
\ref{fig:run6}f).

Third, if $t_{\rm coll}$ is $\lesssim30\:$Myr, then the collision time
is shorter or similar to estimates of GMC lifetimes due to feedback
\citep{WM97,Matzner02,KMM06}. This changes the nature of the life
cycle of GMCs, since they are continually changing and replenishing
their gas content by collisions with other GMCs at rates similar to
those at which they are losing gas by energy and momentum injection
from massive stars. Thus the ``lifetime'' of GMCs no longer has its
traditional meaning, i.e., the time before the GMC is dispersed to
theq atomic or ionized phase of the ISM, since the clock is
essentially ``reset'' after every major collision.

Fourth, GMC collisions may induce star formation, i.e., by compressing
gas that is already on the verge of gravitational instability in the
pre-collision GMCs. The compressed region, with enhanced turbulence
and large velocity gradients, may have its rates of ambipolar
diffusion (e.g., \cite{Mouschovias91,Kunz09}) and/or reconnection
diffusion \citep{Lazarian99,Eyink11} boosted, thus helping remove
magnetic flux and triggering gravitational instabilty and then star
cluster formation. Shear-driven GMC collisions is a process that
naturally connects the kiloparsec scales of galactic orbital dynamics
to the parsec scales of star cluster formation. There have been
numerous claims of GMC collisions being the triggering agents of star
cluster formation: e.g., Westerlund 2 \citep{Furukawa09,Ohama10}, M20
\citep{Torii11}, Cygnus OB 7 \citep{Dobashi14}, and N159 West
\citep{Fukui15}.  GMC collisions are also a natural mechanism for
creating the large observed dispersion in the star formation
efficiencies in GMCs (e.g., \cite{T00,Lee16}).




\section{Conclusions}\label{S:conclusions}

We have simulated flat rotation curve galactic gas disks and
investigated the dynamical evolution of their interstellar media and
GMCs. Based on PDR models, we incorporated a self-consistent treatment
of the ISM physics of the atomic to molecular phase transition under
the influence of various background FUV radiation fields and CR
ionization rates. We then explored the effects of different FUV
intensities, including a model with a radial gradient designed to
mimic the Milky Way galaxy. The effects of cosmic rays, including
radial gradients in their heating and ionization rates, were also
investigated. The final simulation in this series achieved 4 pc
resolution across the $\sim20$ kpc global disk diameter, with heating
and cooling followed down to $\sim10$ K temperatures. The galaxies
were evolved for 300 Myr, i.e., long enough for ISM and GMC properties
to achieve a quasi steady state. We found that during the period of
simulations, the initially marginally stable ISM fragmented into GMCs
due to radiative cooling and gravitational instability. GMC properties
then became influenced by frequent interactions, including collisions.

We examined global ISM properties, including the mass surface density
structures, thermal pressures, and the evolution and spatial
distribution of GMC mass fractions, as well as the GMC mass traced by
CO with $n_{\rm CO}/n_\HH > 10^{-5} {\rm cm^{-3}}$. Our results show
that strong FUV fields can suppress global GMC formation, especially
in the outer regions where gas is more diffuse. Our fiducial model has
a radial profile of molecular mass fraction that is quite similar to
that observed in the Milky Way out to $\sim6\:$kpc by
\citet{Koda16}. The lack of the additional physics of localized star
formation feedback is likely the main cause of the higher molecular
fractions seen in our simulations at larger radii.

We also analyzed GMC properties, including distribution functions of
mass, mass surface density, effective radii, vertical locations,
virial parameters and specific angular momenta. Though we only
incorporated diffuse FUV heating and CR ionization, many observed GMC
properties are approximately reproduced, including the mass spectrum,
mass surface density distribution, and angular momentum distribution,
which shows that retrograde clouds are relatively common. The
distribution of virial parameters shows a large fraction of
gravitationally bound clouds, along with a population of super-virial
clouds potentially caused by strong and frequent cloud-cloud
interactions. We notice that these GMC properties are relatively
insensitive to our different implementations of diffuse FUV and CR
heating, implying rather the important role of GMC
interactions. 

Finally GMC collisions, which may be a means of triggering star
cluster formation, were investigated and compared with analytic models
of galactic shear-induced collisions. Our results show an approximate
agreement with the models, and insensitivity to parameters related to
the diffuse heating. The average collision timescale is relatively
short, i.e., $\sim0.1$ to 0.2 times the local orbital time. Such
frequent GMC collisions may be an important driver of turbulence in
the clouds, potentially explaining their observed structural,
kinematic and dynamical properties.

The simulations so far presented lack several important pieces of
physics, especially magnetic fields and localized star formation
feedback. These aspects, along with the effects of different galactic
dynamical environments, including spiral arms, will be studied in
future papers in this series.

\begin{ack}
Computations described in this work were performed
using the publicly available Enzo code (http://enzo-project.org). This
research also made use of the yt-project (http://yt-project. org/), a
toolkit for analyzing and visualizing quantitative data
\citet{Turk11}. These are products of collaborative efforts of many
independent scientists from numerous institutions around the
world. Their commitment to open science has helped make this work
possible. The authors acknowledge University of Florida Research
Computing (www.rc.ufl.edu) for providing computational resources and
support that have contributed to the research results reported in this
publication.
\end{ack}

\end{document}